\documentclass[publish]{smj}\usepackage[]{graphicx}\usepackage[]{color}
\makeatletter
\def\maxwidth{ %
  \ifdim\Gin@nat@width>\linewidth
    \linewidth
  \else
    \Gin@nat@width
  \fi
}
\makeatother

\definecolor{fgcolor}{rgb}{0.345, 0.345, 0.345}

\usepackage{framed}
\makeatletter
 {\par\unskip\endMakeFramed%
 \at@end@of@kframe}
\makeatother

\definecolor{shadecolor}{rgb}{.97, .97, .97}
\definecolor{messagecolor}{rgb}{0, 0, 0}
\definecolor{warningcolor}{rgb}{1, 0, 1}
\definecolor{errorcolor}{rgb}{1, 0, 0}
\newenvironment{knitrout}{}{} 

\usepackage{alltt}

\usepackage{amsfonts}
\usepackage{float}
\usepackage[section]{placeins}
\usepackage{bm}
\DeclareRobustCommand{\rchi}{{\mathpalette\irchi\relax}}
\newcommand{\irchi}[2]{\raisebox{\depth}{$#1\chi$}}

\usepackage[english, vlined, ruled]{algorithm2e}
\SetAlCapSty{}

\usepackage{booktabs}
\usepackage{tabularx}
\newcolumntype{C}{>{\centering\arraybackslash}X}

\Author{Eduardo E. Ribeiro Jr\Affil{1},
  Walmes M. Zeviani\Affil{2},
  Wagner H. Bonat\Affil{2},
  Clarice G. B. Dem\'{e}trio\Affil{1}
  and John Hinde\Affil{3}
}
\AuthorRunning{Ribeiro Jr \textrm{et al.}}

\Affiliations{
\item Department of Exact Sciences,
  University of S\~{a}o Paulo - ESALQ,
  Piracicaba,
  SP, Brazil
\item Department of Statistics,
  Paran\'{a} Federal University,
  Curitiba,
  PR, Brazil
\item School of Mathematics, Statistics and Applied Mathematics,
  National University of Ireland,
  Galway,
  Galway, Ireland
}

\CorrAddress{Eduardo E. Ribeiro Jr,
             Department of Exact Sciences,
             Luiz de Queiroz College of Agriculture - ESALQ,
             University of S\~{a}o Paulo - USP,
             Piracicaba, S\~{a}o Paulo,
             P\'{a}dua Dias, 11, Avenue, CEP--13.418-900
             Brazil}
\CorrEmail{jreduardo@usp.br}
\CorrPhone{(+55)\;41\;9\;8711\;9034}
\CorrFax{(+55)\;19\;3447\;6021}

\Title{Reparametrization of COM-Poisson Regression Models with
  Applications in the Analysis of Experimental Data}
\TitleRunning{Reparametrization of COM-Poisson Regression Models}

\Abstract{ In the analysis of count data often the equidispersion
  assumption is not suitable, hence the Poisson regression model is
  inappropriate.  As a generalization of the Poisson distribution the
  COM-Poisson distribution can deal with under-, equi- and overdispersed
  count data.  It is a member of the exponential family of distributions
  and has the Poisson and geometric distributions as special cases, as
  well as the Bernoulli distribution as a limiting case.  In spite of
  the nice properties of the COM-Poisson distribution, its location
  parameter does not correspond to the expectation, which complicates
  the interpretation of regression models specified using this
  distribution. In this paper, we propose a straightforward
  reparametrization of the COM-Poisson distribution based on an
  approximation to the expectation of this distribution. The main
  advantage of our new parametrization is the straightforward
  interpretation of the regression coefficients in terms of the
  expectation of the count response variable, as usual in the context of
  generalized linear models. Furthermore, the estimation and inference
  for the new COM-Poisson regression model can be done based on the
  likelihood paradigm. We carried out simulation studies to verify the
  finite sample properties of the maximum likelihood estimators. The
  results from our simulation study show that the maximum likeli-hood
  estimators are unbiased and consistent for both regression and
  dispersion parameters. We observed that the empirical correlation
  between the regression and dispersion parameter estimators is close to
  zero, which suggests that these parameters are orthogonal. We
  illustrate the application of the proposed model through the analysis
  of three data sets with over-, under- and equidispersed count data.
  The study of distribution properties through a consideration of
  dispersion, zero-inflated and heavy tail indices, together with the
  results of data analysis show the flexibility over standard
  approaches. Therefore, we encourage the application of the new
  parametrization for the analysis of count data in the context of
  COM-Poisson regression models.  The computational routines for fitting
  the original and new version of the COM-Poisson regression model and
  the analyzed data sets are available in the supplementary material.
}


\IfFileExists{upquote.sty}{\usepackage{upquote}}{}
\begin{document}

\maketitle

\section{Introduction}
\label{introduction}

Count data are random variables that assume non-negative integer values
and represent the number of times an event occurs in the observation
period. This kind of data is common in crop sciences, such as the number
of grains produced by a plant, number of fruits produced by a tree,
number of insects captured by a trap, to cite but a few.  Since the
seminal paper of \citet{Nelder1972} where the class of the generalized
linear models (GLMs) was introduced, the analysis of count data often
employs the Poisson regression model. This model provides a suitable
strategy for the analysis of count data and an efficient Newton scoring
algorithm can be used for fitting the model.

In spite of the advantages of the Poisson regression model, the Poisson
distribution has only one parameter, which represents both the
expectation and variance of the count random variable.  This restriction
on the relationship between the expectation and variance induced by the
Poisson distribution is referred as equidispersion.  However, in
practical data analysis such a restriction can be unsuitable, since the
observed data can present  variance both smaller or larger than the
expectation, leading to the cases of under and overdispersion,
respectively.  The main problem of the application of the Poisson
regression model to non-equidispersed count data is that the standard
errors associated with the regression coefficients are inconsistently
estimated, which in turn can lead to misleading
inferences~\citep{Winkelmann1995, Bonat2017}.

In practice, overdispersion is largely reported in the literature and
may occur due to the absence of relevant covariates, heterogeneity of
sampling units, different observational periods/regions not considered
in the analysis, and excess of zeros~\citep{Hinde1998}.  The case of
underdispersion is less report in the literature, however, it has been
of increasing interest in the statistical community. The processes that
reduce the variability are not as well-known as those leading to extra
variability.  For this reason, there are few approaches to deal with
underdispersed count data. The explanatory mechanisms leading to
underdispersion may be related to the underlying stochastic process
generating the count data.  When the time between events is not
exponentially distributed, the number of events can be over or
underdispersed, this process motivated the class of duration dependence
models~\citep{Winkelmann1995}.  Another possible explanation of
underdispersion is when the responses correspond to order statistics of
component observations, such as maxima of Poisson distributed
counts~\citep{Steutel1989}.

The strategies for constructing alternative count distributions are
related with the causes of the non-equidispersion. Specfically for the
overdispersion case Poisson mixture models are widely applied.  One
popular example of this approach is the negative-binomial model, where
the expectation of the Poisson distribution is assumed to be gamma
distributed.  However, other distributions can be used to represent the
random variation.  For example the Poisson-Tweedie
model~\citep{Bonat2017} and its special cases as the Poisson
inverse-Gaussian and Neyman-Type A assume that the random effects are
Tweedie, inverse Gaussian and Poisson distributed, respectively.  The
Gamma-Count distribution assumes a gamma distribution for the time
between events, thus it can deal with underdispersed as well as
overdispersed count data~\citep{Zeviani2014}. The COM-Poisson
distribution is obtained by a generalization of the Poisson distribution
allowing for a non-linear decrease in the ratios of successive
probabilities~\citep{Shmueli2005}.

The COM-Poisson distribution is a member of the exponential family and
it has the Poisson and geometric distributions as special cases, as well
as the Bernoulli distribution as a limiting case. It can deal with both
under and overdispersed count data.  Some recently applications of the
COM-Poisson distribution include \citet{Lord2010} for the analysis of
traffic crash data, \citet{Sellers2010} for the modelling of airfreight
breakage and book purchases, and \citet{Huang2017} to the analysis of
attendance data, takeover birds and cotton boll counts.  The main
disadvantage of the COM-Poisson regression model as presented
in~\citet{Sellers2010} is that its location parameter does not
correspond to the expectation of the distribution, which complicates the
interpretation of regression models and means that they are not
comparable with standard approaches such as the Poisson and negative
binomial regression models. \citet{Huang2017} proposed a
mean-parametrization of the COM-Poisson distribution in order to avoid
such an issue.  In this approach the mean parameter is obtained by
solving an non-linear equation defined as an infinite sum.
Consequently, it is computationally demanding and liable to numerical
problems.

The main goal of this article is to propose a novel COM-Poisson
parametrization based on the mean approximation presented by
\citet{Shmueli2005}. In this parametrization, the probability mass
function is written in terms of $\mu$ and $\phi$, where $\mu$ is the
expectation and $\phi$ is a dispersion parameter. In contrast to the
original parametrization, the proposed parametrization leads to
regression coefficients directly associated with the expectation of the
response variable, as usual in the context of generalized linear models.
Consequently, the obtained regression coefficients are comparable with
the ones obtained by standard approaches, such as the Poisson and
negative binomial regression models.  Furthermore, our novel COM-Poisson
parametrization is simpler than the strategy proposed
by~\cite{Huang2017}, since it does not require any numerical method for
solving non-linear equations, and we show the attractive properties like
the orthogonality between dispersion and regression parameters and
consistency and asymptotic normality of the maximum likelihood
estimators are retained.

This paper is organized as follows. In \autoref{background} we present
the COM-Poisson distribution and the strategy proposed
by~\cite{Huang2017}.  The proposed reparametrization, assessment of
moment approximations, and study of distribution properties are
considered in the~\autoref{reparametrization}. In
the~\autoref{estimation-and-inference} we present estimation and
inference for the novel COM-Poisson regression model based on the
likelihood paradigm. The properties of the maximum likelihood estimators
and the orthogonality property are assessed
in~\autoref{simulation-study} through simulation studies.  We illustrate
the application of the new COM-Poisson regression model through the
analysis of three data sets. We provide an \texttt{R} implementation of
the COM-Poisson and reparameterized COM-Poisson regression models as
well as the analyzed data sets in the supplementary
material.\footnote{Available on
  \texttt{\url{http://www.leg.ufpr.br/~eduardojr/papercompanions}}
  \label{papercompanion}.}.

\section{Background}
\label{background}

The COM-Poisson distribution generalizes the Poisson distribution in
terms of the ratio between the probabilities of two consecutive events
by adding an extra dispersion parameter~\citep{Sellers2010}.  Let $Y$ be
a COM-Poisson random variable, then
$$\frac{\Pr(Y=y-1)}{\Pr(Y=y)} = \frac{y^\nu}{\lambda}$$ while for the
Poisson distribution this ratio is $\frac{y}{\lambda}$ corresponding to
$\nu=1$.  It allows the COM-Poisson distribution deals with
non-equidispersed count data.  The probability mass function of the
COM-Poisson distribution is given by
\begin{equation}
  \label{eqn:pmf-cmp}
  \Pr(Y=y \mid \lambda, \nu) = \frac{\lambda^y}{(y!)^\nu Z(\lambda, \nu)},
  \qquad y = 0, 1, 2, \ldots,
\end{equation}
where $\lambda > 0$, $\nu \geq 0$ and $Z(\lambda, \nu) =
\sum_{j=0}^\infty \frac{\lambda^j}{(j!)^\nu}$ is a normalizing
constant that depends on both parameters.

The $Z(\lambda, \nu)$ series diverges theoretically only when $\nu=0$
and $\lambda \geq 1$, but numerically for small values of $\nu$ combined
with large values of $\lambda$, the sum is so huge it causes
overflow. \autoref{tab:convergenceZ} shows the values of the normalizing
constants using one thousand increments, that is,
$\sum_{j=0}^{1000}\lambda^j/(j!)^\nu$ for different values of $\lambda$
and $\phi$.

\begin{table}[ht]
\centering
\caption{Values for $Z(\lambda, \nu)$ constant (numerically computed) for values of $\lambda$ (0.5 to 50) and $\phi$ (0 to 1)}
\label{tab:convergenceZ}
\begingroup\small
\begin{tabularx}{\textwidth}{C|CCCCCC}
  \toprule

  & \multicolumn{6}{c}{$\bm{\lambda}$} \\
 $\bm{\nu}$ & 0.5 & 1 & 5 & 10 & 30 & 50 \\
 \midrule
0 &  2.00     & divergent$^{*\,\,\,}$ & divergent$^{*\,\,\,}$ & divergent$^{*\,\,\,}$ & divergent$^{*\,\,\,}$ & divergent$^{*\,\,\,}$ \\ 
  0.1 &  1.92     &  7.64     &       divergent$^{**}$ &       divergent$^{**}$ &       divergent$^{**}$ &       divergent$^{**}$ \\ 
  0.2 &  1.86     &  5.25     & 3.17e+273 &       divergent$^{**}$ &       divergent$^{**}$ &       divergent$^{**}$ \\ 
  0.3 &  1.81     &  4.32     &  1.60e+29 & 2.54e+282 &       divergent$^{**}$ &       divergent$^{**}$ \\ 
  0.4 &  1.77     &  3.80     &  4.71e+10 &  1.33e+56 &       divergent$^{**}$ &       divergent$^{**}$ \\ 
  0.5 &  1.74     &  3.47     &  1.34e+06 &  3.67e+22 & 3.32e+196 &       divergent$^{**}$ \\ 
  0.6 &  1.72     &  3.23     &  2.05e+04 &  4.99e+12 &  1.73e+76 & 4.63e+177 \\ 
  0.7 &  1.70     &  3.06     &  2.37e+03 &  3.69e+08 &  4.93e+39 &  6.93e+81 \\ 
  0.8 &  1.68     &  2.92     &  6.49e+02 &  2.70e+06 &  5.09e+24 &  3.43e+46 \\ 
  0.9 &  1.66     &  2.81     &  2.74e+02 &  1.47e+05 &  1.80e+17 &  2.19e+30 \\ 
  1 &  1.65     &  2.72     &  1.48e+02 &  2.20e+04 &  1.07e+13 &  5.18e+21 \\ 
   \bottomrule

\end{tabularx}
\endgroup

\footnotesize \raggedright
divergent$^{*}$ is a mathematically divergent series; and
divergent$^{**}$ a numerically divergent series.
\end{table}

In the first line of \autoref{tab:convergenceZ} we have mathematically
divergent series, because $\sum_{j=0}^\infty\lambda^j$ is divergent when
$\lambda \geq 1$.  In other cases the series diverges numerically, due
to the computational storage limitation.  For both forms of divergence
it is impossible to compute probabilities, therefore, this acts as a
restriction on the parameter space.

An undesirable feature of the COM-Poisson distribution is that the
moments cannot be obtained in closed form. \citet{Shmueli2005} and
\citet{Sellers2010} using an asymptotic approximation for
$Z(\lambda,\nu)$, showed that the expectation and variance of the
COM-Poisson distribution can be approximated by
\begin{equation}
  \label{eqn:mean-approx}
  \text{E}(Y) \approx \lambda^{1/\nu} - \frac{\nu - 1}{2\nu} \qquad
  \textrm{and} \qquad
  \textrm{Var}(Y) \approx \frac{\lambda^{1/\nu}}{\nu}\,,
\end{equation} which is particularly accurate for  $\nu \leq 1$ or
$\lambda > 10$. The authors also argue that the mean-variance
relationship can be approximate by
$\frac{1}{\nu}\text{E}(Y)$. In~\autoref{reparametrization}, we assess
the accuracy of these approximations.

The COM-Poisson regression model was proposed by~\citet{Sellers2010},
using the original parametrization. In this case, the COM-Poisson
regression model is $\log(\lambda_i) = \bm{x}_i^\top \bm{\beta}$ and the
relationship between E$(Y_i)$ and $\bm{x}_i$ is modelled indirectly.
\citet{Huang2017} shows how to model directly the expectation of the
COM-Poisson distribution in a suitable parametrization. In the
\autoref{eqn:pmf-cmp}, \Citeauthor{Huang2017} proposes that the
parameter $\lambda$ as a function of $\mu$ and $\nu$, is given by the
solution to
\begin{equation*}
  \sum_{j=0}^{\infty} (j - \mu) \frac{\lambda^j}{(y!)^\nu} = 0\,.
\end{equation*}
Thus the mean-parametrized COM-Poisson regression model is $\log(\mu_i)
= \bm{x}_i^\top \bm{\beta}$. In this article, we propose an alternative
mean-parametrization of the COM-Poisson distribution in order to avoid the
limitations of the original parametrization and the numerical complexity
of the \Citeauthor{Huang2017}'s approach.

\section{Reparametrized COM-Poisson regression model}
\label{reparametrization}

The proposed reparametrization of COM-Poisson models is based on
the mean approximation (\autoref{eqn:mean-approx}). We introduced a new
parameter $\mu$, using this approximation,
\begin{equation}
  \label{eqn:repar-cmp}
  \mu = h(\lambda, \nu) = \lambda^{1/\nu} - \frac{\nu - 1}{2\nu}
  \quad \Rightarrow \quad
  \lambda = h^{-1}(\mu, \nu) = \left (\mu +
    \frac{(\nu - 1)}{2\nu} \right )^\nu.
\end{equation}
The dispersion parameter is taken on the log scale for computational
convenience, thus $\phi = \log(\nu)$, $\phi \in \mathbb{R}$. The
interpretation of $\phi$ is the same as the $\nu$, but on another
scale. For $\phi < 0$ and $\phi > 0$ we have the overdispersed and
underdispersed cases, respectively.  When $\phi=0$ we have Poisson
distribution as a special case.

In order to assess the accuracy of the moment approximations
(\autoref{eqn:mean-approx}), \autoref{fig:approx-plot} presents the
quadratic errors for (a) expectation  and (b) variance.  The
quadratic errors were obtained by $[\mu - \text{E}(Y)]^2$ for the
expectation and by $[ \mu \exp(-\phi) - \text{Var}(Y)]^2$ for the
variance. In both cases $\text{E}(Y)$ and Var$(Y)$ were computed
numerically. The dotted lines represent the border between the regions
$\nu \leq 1$ and $\lambda > 10^\nu$, in the $\mu$ and $\phi$ scale.

\begin{knitrout}
\definecolor{shadecolor}{rgb}{0.969, 0.969, 0.969}\color{fgcolor}\begin{figure}[!htp]

{\centering \includegraphics[width=.8\textwidth]{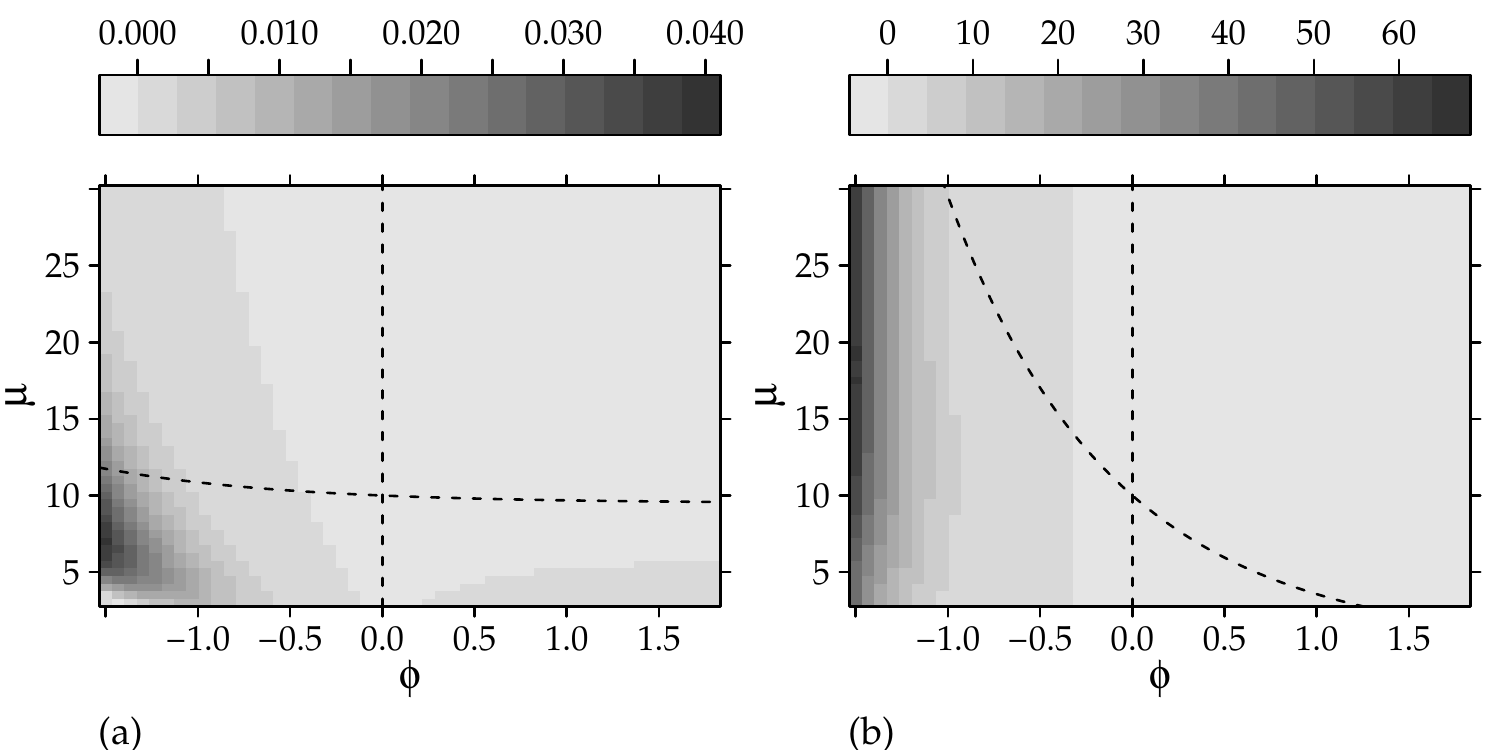} 

}

\caption[Quadratic errors for the approximation of the (a) expectation and (b) variance]{Quadratic errors for the approximation of the (a) expectation and (b) variance. Dotted lines represent the restriction for suitable approximations given by \cite{Shmueli2005}.}\label{fig:approx-plot}
\end{figure}

\end{knitrout}

The results in \autoref{fig:approx-plot} show that the mean
approximation is accurate, the largest quadratic error is
0.038 for the parameter values evaluated. For the
variance approximation, the largest quadratic error was
63.903 and it occurs for negative values of
$\phi$. Interestingly, the errors are larger for negative values of
$\phi$ and present no clear relation with $\mu$, as opposed to the
regions gives by \citet{Shmueli2005} ($\phi \leq 0$ and
$\mu > 10 - \frac{\exp(\phi) - 1}{2\exp(\phi)}$).

The results presented in \autoref{fig:approx-plot}(a) support the
proposed reparametrization. Replacing $\lambda$ and $\nu$ as function of
$\mu$ and $\phi$ in \autoref{eqn:pmf-cmp}, the reparametrized
distribution takes the form
\begin{equation}
  \label{eqn:pmf-cmpmu}
  \Pr(Y=y \mid \mu, \phi) =
  \left ( \mu +\frac{ e^\phi-1}{2e^\phi} \right )^{ye^\phi}
  \frac{(y!)^{-e^\phi}}{Z(\mu, \phi)},
  \qquad y = 0, 1, 2, \ldots\,,
\end{equation} where $\mu > 0$. We denote this distribution as
COM-Poisson$_\mu$. In \autoref{fig:pmf-cmp}, we show the shapes of
COM-Poisson$_\mu$ distribution.

\begin{knitrout}
\definecolor{shadecolor}{rgb}{0.969, 0.969, 0.969}\color{fgcolor}\begin{figure}[!htp]

{\centering \includegraphics[width=1\textwidth]{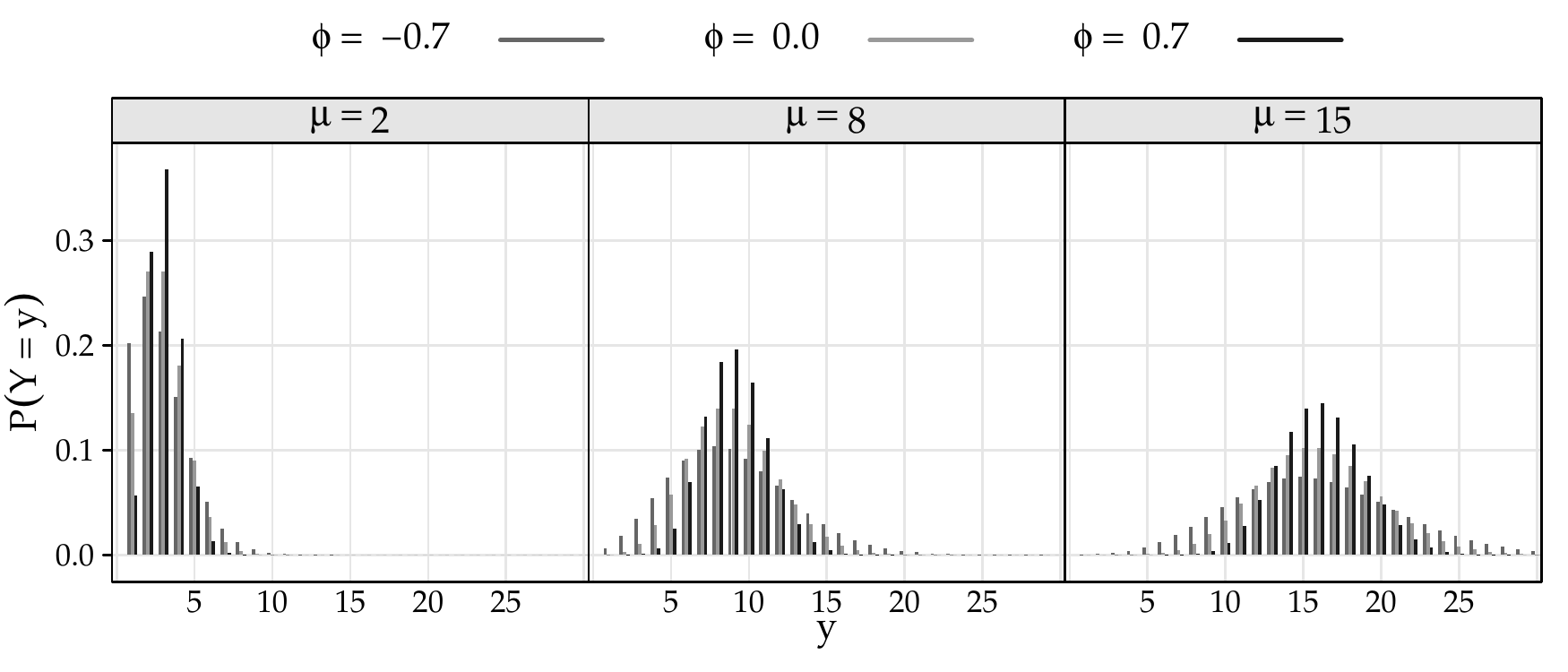} 

}

\caption[Shapes of the COM-Poisson distribution for different parameter values]{Shapes of the COM-Poisson distribution for different parameter values.}\label{fig:pmf-cmp}
\end{figure}

\end{knitrout}

In order to explore the flexibility of the COM-Poisson model to deal
with real count data, we compute indexes for dispersion (DI),
zero-inflation (ZI) and heavy-tail (HI), which are respectively given by
\begin{equation*}
\text{DI} = \frac{\text{Var}(Y)}{\text{E}(Y)}, \quad
\text{ZI} = 1 + \frac{\log \Pr(Y = 0)}{\text{E}(Y)}
  \quad \text{and} \quad
\text{HT} = \frac{\Pr(Y=y+1)}{\Pr(Y=y)}\quad \text{for} \quad y \to
\infty.
\end{equation*}
These indexes are defined in relation to the Poisson distribution. Thus,
the dispersion index indicates overdispersion for $\text{DI} > 1$,
underdispersion for $\text{DI} < 1$ and equidispersion for
$\text{DI} = 1$. The zero-inflation index indicates zero-inflation for
$\text{ZI} > 0$, zero-deflation for $\text{ZI} < 0$ and no excess of
zeros for $\text{ZI} = 0$. Finally, heavy-tail index indicates a
heavy-tail distribution for $\text{HT} \to 1$ when $y \to \infty$.

These indexes are discussed by \citet{Bonat2017} to study the
flexibility of Poisson-Tweedie distribution, and \citet{Puig2006} to
describe count distributions. Regarding the COM-Poisson$_\mu$
distribution, in \autoref{fig:indexes-plot} we present the relationship
between (a) mean and variance, (b--c) the dispersion and zero-inflation
indexes for different values of $\mu$ and $\phi$, and (d) heavy-tail
index for $\mu=25$ and different values of $y$ and $\phi$.

\begin{knitrout}
\definecolor{shadecolor}{rgb}{0.969, 0.969, 0.969}\color{fgcolor}\begin{figure}[!htp]

{\centering \includegraphics[width=1\textwidth]{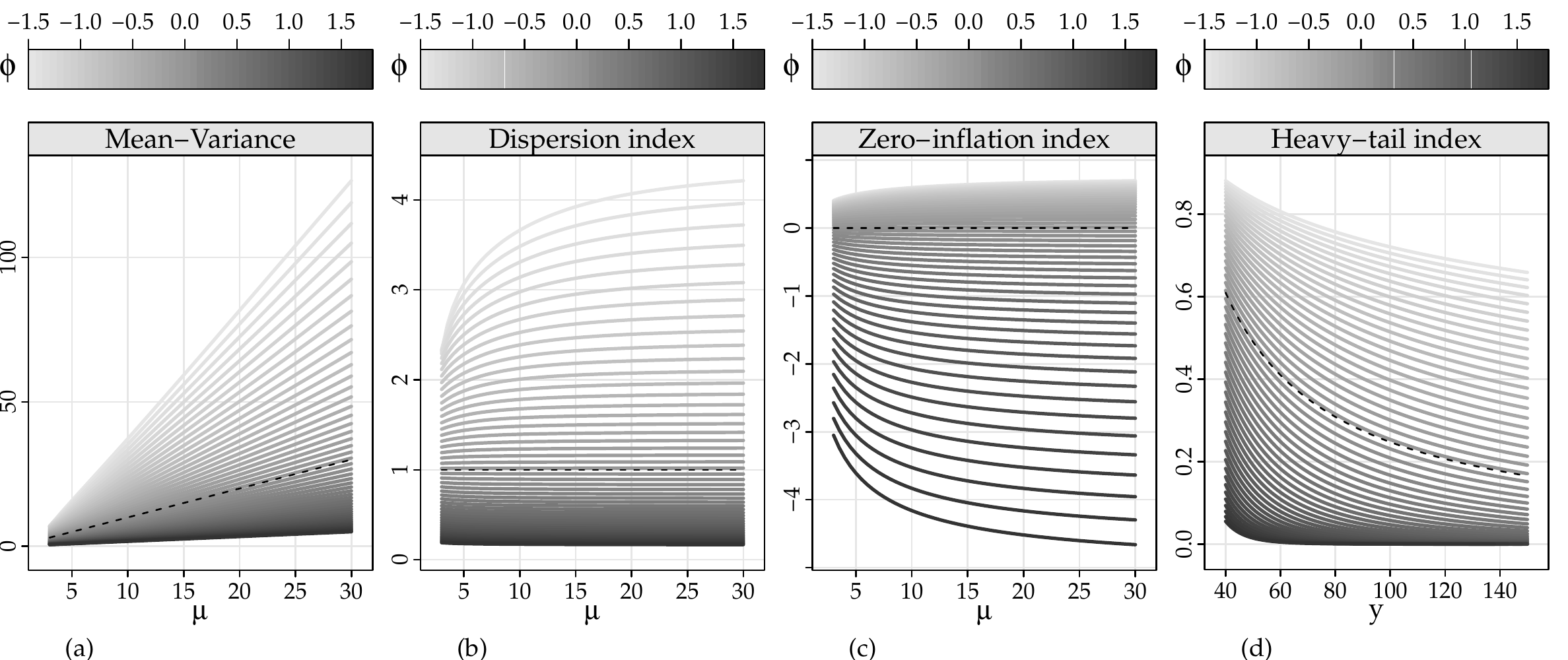} 

}

\caption[Indexes for COM-Poisson distribution]{Indexes for COM-Poisson distribution. (a) Mean and variance relationship, (b--d) dispersion, zero-inflation and heavy-tail indexes for different parameter values. Dotted lines represents the Poisson special case.}\label{fig:indexes-plot}
\end{figure}

\end{knitrout}

\autoref{fig:indexes-plot} shows that the indexes are slightly dependent
on the expected values and tend to stabilize for large values of
$\mu$. Consequently, the mean and variance relationship
\autoref{fig:indexes-plot}(a) is proportional to the dispersion
parameter $\phi$. In terms of moments, this leads to a specification
indistinguishable from the quasi-Poisson regression model. The
dispersion indexes in \autoref{fig:indexes-plot}(b) show that the
distribution is suitable to deal to dispersed counts, of course. For the
parameter values evaluated the largest DI was 4.213 and smallest
was 0.168. \autoref{fig:indexes-plot}(c) shows the COM-Poisson
can handle a limited amount of zero-inflation, in cases of
overdispersion ($\phi < 0$). On the other hand, for $\phi > 0$
(underdispersion) this distribution is suitable to deal with
zero-deflated counts. Heavy-tail indexes in
\autoref{fig:indexes-plot}(d) indicate the distribution is in general a
light-tailed distribution, i.e. $HT \to 0$ for $y \to \infty$.

\section{Estimation and Inference}
\label{estimation-and-inference}

In this section we describe the estimation and inference for the
two forms of the COM-Poisson regression model based on the maximum
likelihood method. The log-likelihood function for a set of independent
observations $y_i$, $i=1,2,\ldots,n$ from the COM-Poisson$_\mu$
distribution has the following form,
\begin{equation}
  \label{eqn:ll-rcmp}
  \ell = \ell(\bm{\beta}, \phi \mid \bm{y}) =
  e^\phi \left [
    \sum_{i=1}^n y_i
    \log \left( \mu_i + \frac{e^\phi-1}{2e^\phi} \right ) -
    \sum_{i=1}^n \log(y_i!) \right ] -
  \sum_{i=1}^n \log(Z(\mu_i, \phi)),
\end{equation}
where $\mu_i = \exp(\bm{x}_i^\top\bm{\beta})$, with
$\bm{x}_i^\top = (x_{i1},\, x_{i2},\, \ldots,\, x_{ip})$ is a vector of
known covariates for the $i$-th observation, and $(\bm{\beta},\, \phi)
\in \mathbb{R}^{p+1}$. The normalizing constant $Z(\mu_i, \phi)$ is
given by
\begin{equation}
\label{eqn:infseries}
  Z(\mu_i, \phi) = \sum_{j=0}^\infty \left [ \left (
    \mu_i + \frac{e^\phi - 1}{2e^\phi} \right )^{je^\phi}
  \frac{1}{(j!)^{e^\phi}} \right ].
\end{equation}
The evaluation of the log-likelihood function for each observation
involves the computation of the infinite series
(\autoref{eqn:infseries}).  Thus, the fitting procedure is computationally
expensive for regions of the parameter space where the convergence of
the infinite sum is slow.

Parameter estimation requires the numerical maximization of
\autoref{eqn:ll-rcmp}. Since the derivatives of $\ell$ cannot be
obtained in closed forms, we adopted the \texttt{BFGS}
algorithm~\citep{Nocedal1995} as implemented in the function
\texttt{optim()} for the statistical software \texttt{R}
\citep{Rcore2017}. Standard errors for the regression coefficients are
obtained based on the observed information matrix
$\mathcal{I}(\bm{\theta})$, where
$\mathcal{I}(\bm{\theta}) = -\mathcal{H}(\bm{\theta})$ (hessian matrix)
is computed numerically. Confidence intervals for $\hat{\mu}_i$ are
obtained by using the delta method~\citep[p. 89]{Pawitan2001}.

The parameter estimation for the COM-Poisson regression model in the
original parametrization is analogous to the one presented for the
COM-Poisson$_\mu$ distribution, however, it considers
\autoref{eqn:ll-rcmp} in terms of $\lambda$. Even for the standard
COM-Poisson distribution, the dispersion parameter is taken on the log
scale to avoid numerical issues.

In the applications we fitted the quasi-Poisson model
\citep{Wedderburn1974} as a baseline model. This approach is based on a
second-moment assumption that allows more flexibility to the model. In
this case the variance of the response variable is fixed by an
additional parameter $\sigma$, $\textrm{Var}(Y_i)=\sigma \mu_i$. These
models are fitted in the \texttt{R} software using the function
\texttt{glm(..., family = quasipoisson)}.

\section{Simulation study}
\label{simulation-study}

In this section we performed a simulation study to assess the properties
of the maximum likelihood estimators and orthogonality of the
reparametrized model as well as the flexibility of the COM-Poisson
regression model to deal with non-equidispersed count data.

We considered average counts varying from $3$ to $27$ according to a
regression model with a continuous and a categorical covariate. The
continuous covariate~($\bm{x}_1$) was generated as a sequence from $0$
to $1$ and of length equal to the sample size.  Similarly, the categorical
covariate~($\bm{x}_2$) was generated as a sequence of three values each
one repeated $n/3$ times (rounding up when required), where $n$ denotes
the sample size. Thus, the expectation of the COM-Poisson random
variable is given by
$\bm{\mu} = \exp(\beta_0 + \beta_1 \bm{x}_1 + \beta_{21} \bm{x}_{21} +
\beta_{22} \bm{x}_{22})$,
where $\bm{x}_{21}$ and $\bm{x}_{22}$ are dummy representing the levels
of $\bm{x}_2$.  The regression coefficients were fixed at the values,
$\beta_0 = 2$, $\beta_1 = 0.5$, $\beta_{21} = 0.8$ and
$\beta_{22} = -0.8$.

We designed four simulation scenarios by considering different values of
the dispersion parameter $\phi = -1.6, -1.0, 0.0$ and $1.8$.  Thus, we
have strong and moderate overdispersion, equidispersion, and
underdispersion, respectively.  \autoref{fig:justpars} shows the
variation of the average counts (left) and dispersion index (right) for
each value of the dispersion parameter considered in the simulation
study.  These configurations allow us to assess the properties of the
maximum likelihood estimators in extreme situations, such as high counts
and low dispersion, and low counts and high dispersion, but also in the
standard case of equidispersion.

\begin{knitrout}
\definecolor{shadecolor}{rgb}{0.969, 0.969, 0.969}\color{fgcolor}\begin{figure}[!htp]

{\centering \includegraphics[width=1\textwidth]{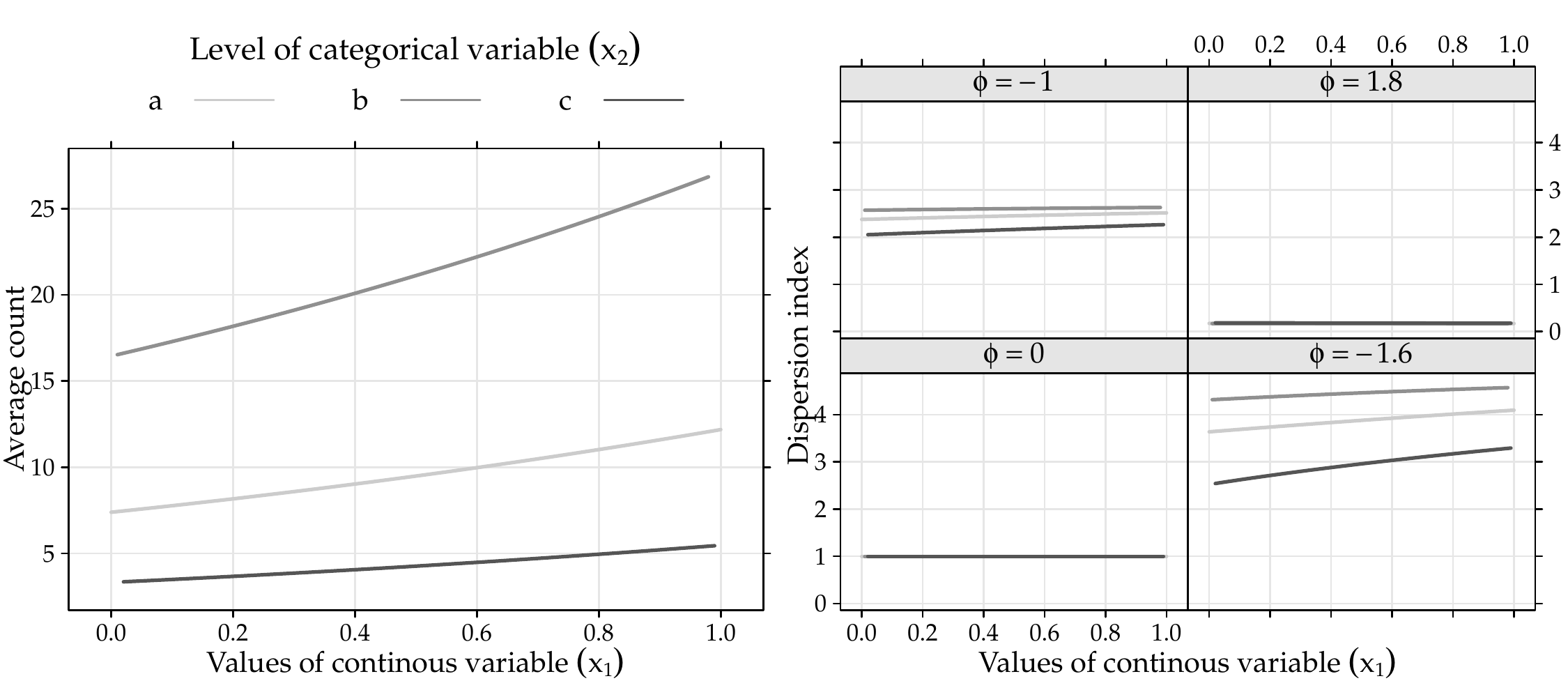} 

}

\caption[Average counts (left) and dispersion indexes (right) for each scenario considered in the simulation study]{Average counts (left) and dispersion indexes (right) for each scenario considered in the simulation study.}\label{fig:justpars}
\end{figure}

\end{knitrout}

In order to check the consistency of the estimators we considered four
different sample sizes: $50$, $100$, $300$ and $1000$; generating $1000$
data sets in each case.  In \autoref{fig:bias-plot}, we show the bias of
the estimators for each simulation scenario (combination between values
of the dispersion parameter and samples sizes) along with the confidence
intervals calculated as average bias plus and minus $\Phi(0.975)$ times
the average standard error. The scales are standardized for each
parameter by dividing the average bias by the average standard error
obtained for the sample of size $50$.

The results in \autoref{fig:bias-plot} show that for all dispersion
levels, both the average bias and standard errors tend to $0$ as the
sample size increases. The estimators for the regression parameters are
unbiased, consistent and their empirical distributions are
symmetric. For the dispersion parameter, the estimator is asymptotically
unbiased; in small samples the parameter is overestimated and the
empirical distribution is slightly right-skewed.

\begin{knitrout}
\definecolor{shadecolor}{rgb}{0.969, 0.969, 0.969}\color{fgcolor}\begin{figure}[!htp]

{\centering \includegraphics[width=1\textwidth]{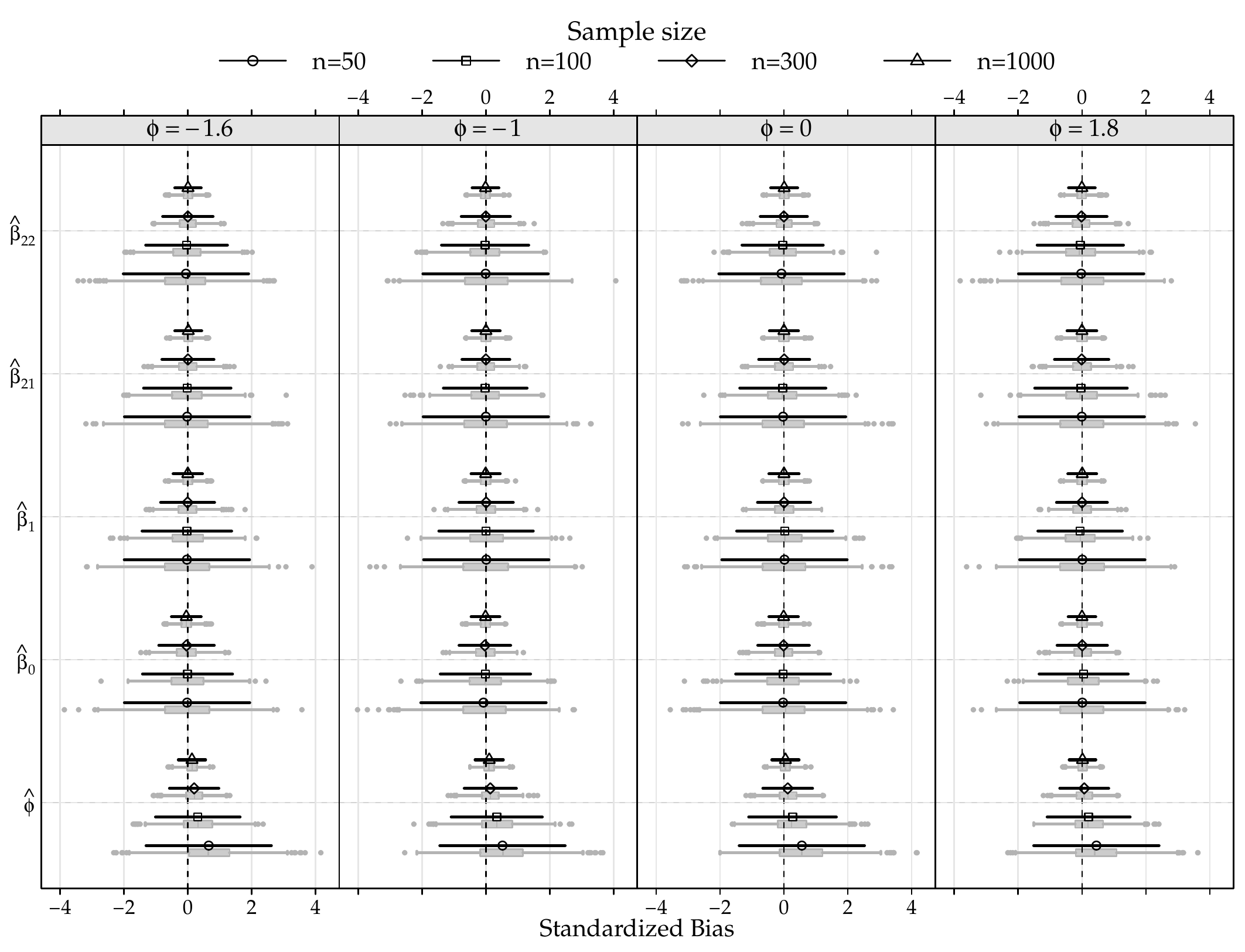} 

}

\caption[Distributions of standardized bias (gray box-plots) and average with confidence intervals (black segments) by different sample sizes and dispersion levels]{Distributions of standardized bias (gray box-plots) and average with confidence intervals (black segments) by different sample sizes and dispersion levels.}\label{fig:bias-plot}
\end{figure}

\end{knitrout}

\autoref{fig:coverage-plot} presents the empirical coverage rate of the
asymptotic confidence intervals. The results show that for the
regression parameters the empirical coverage rates are close to the
nominal level of 95\% for sample sizes greater than $100$ and all
simulation scenarios. For the dispersion parameter the empirical
coverage rates are slightly lower than the nominal level; however, they
become closer to the nominal level for large samples.  The worst
scenario is when we have small sample size and strong overdispersed
counts.

\begin{knitrout}
\definecolor{shadecolor}{rgb}{0.969, 0.969, 0.969}\color{fgcolor}\begin{figure}[!htp]

{\centering \includegraphics[width=1\textwidth]{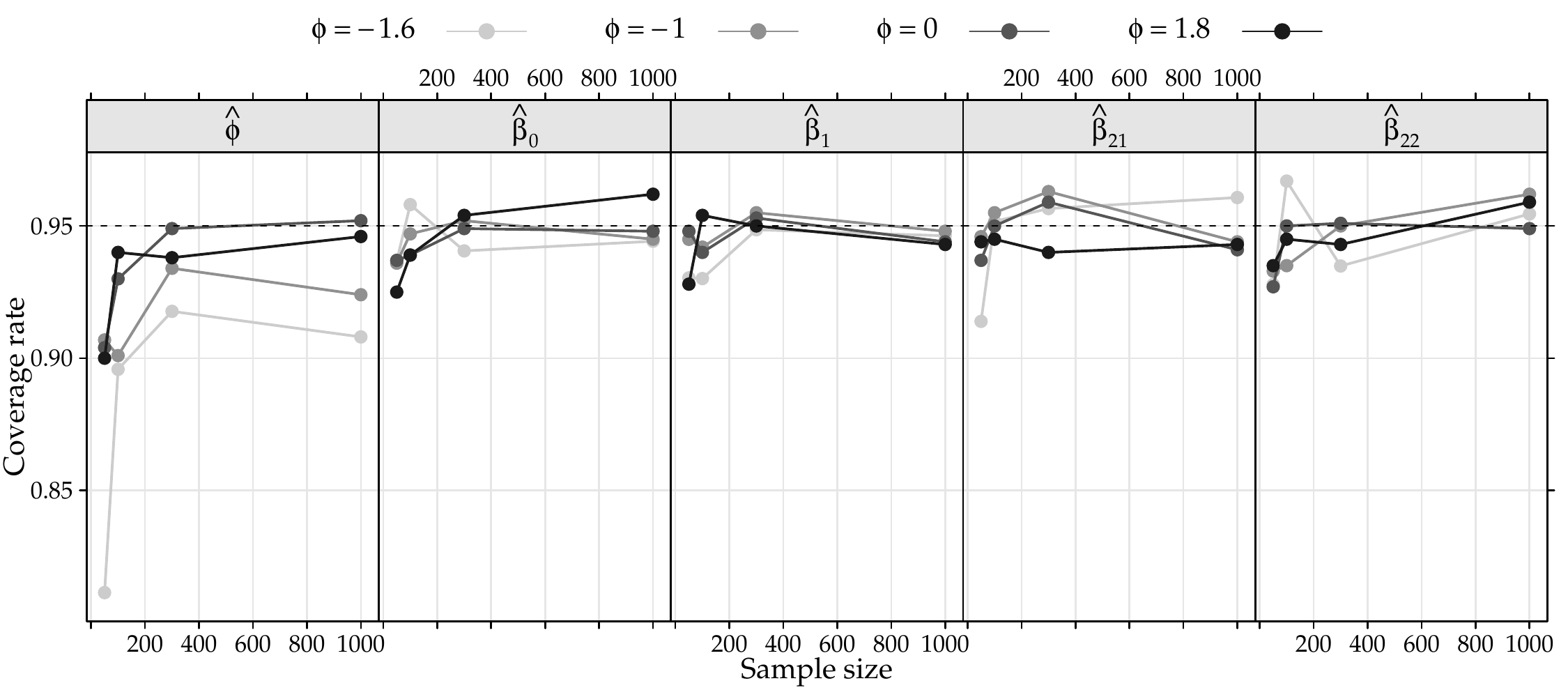} 

}

\caption[Coverage rate based on confidence intervals obtained by quadratic approximation for different sample sizes and dispersion levels]{Coverage rate based on confidence intervals obtained by quadratic approximation for different sample sizes and dispersion levels.}\label{fig:coverage-plot}
\end{figure}

\end{knitrout}

To check the orthogonality property we compute the covariance matrix
between maximum likelihood estimators
$\hat{\bm{\theta}} = (\hat{\bm{\beta}}, \phi)$, obtained from the  observed
information matrix,
Cov$(\hat{\bm{\theta}}) = \mathcal{I}^{-1}(\bm{\theta})$.
\autoref{fig:ortho-plot} shows the covariance between regression and
dispersion parameter estimators for each simulation scenario, on
the correlation scale. The correlations are close to zero in all cases
suggesting the orthogonality property for the reparametrized
model. Interestingly, results in the first panel show that
cov($\hat{\beta}_{22}, \hat{\phi}$) is not very close to zero (values
between $-0.4$ and $0.2$) for strong overdispersion ($\phi = -1.6$).

\begin{knitrout}
\definecolor{shadecolor}{rgb}{0.969, 0.969, 0.969}\color{fgcolor}\begin{figure}[!htp]

{\centering \includegraphics[width=1\textwidth]{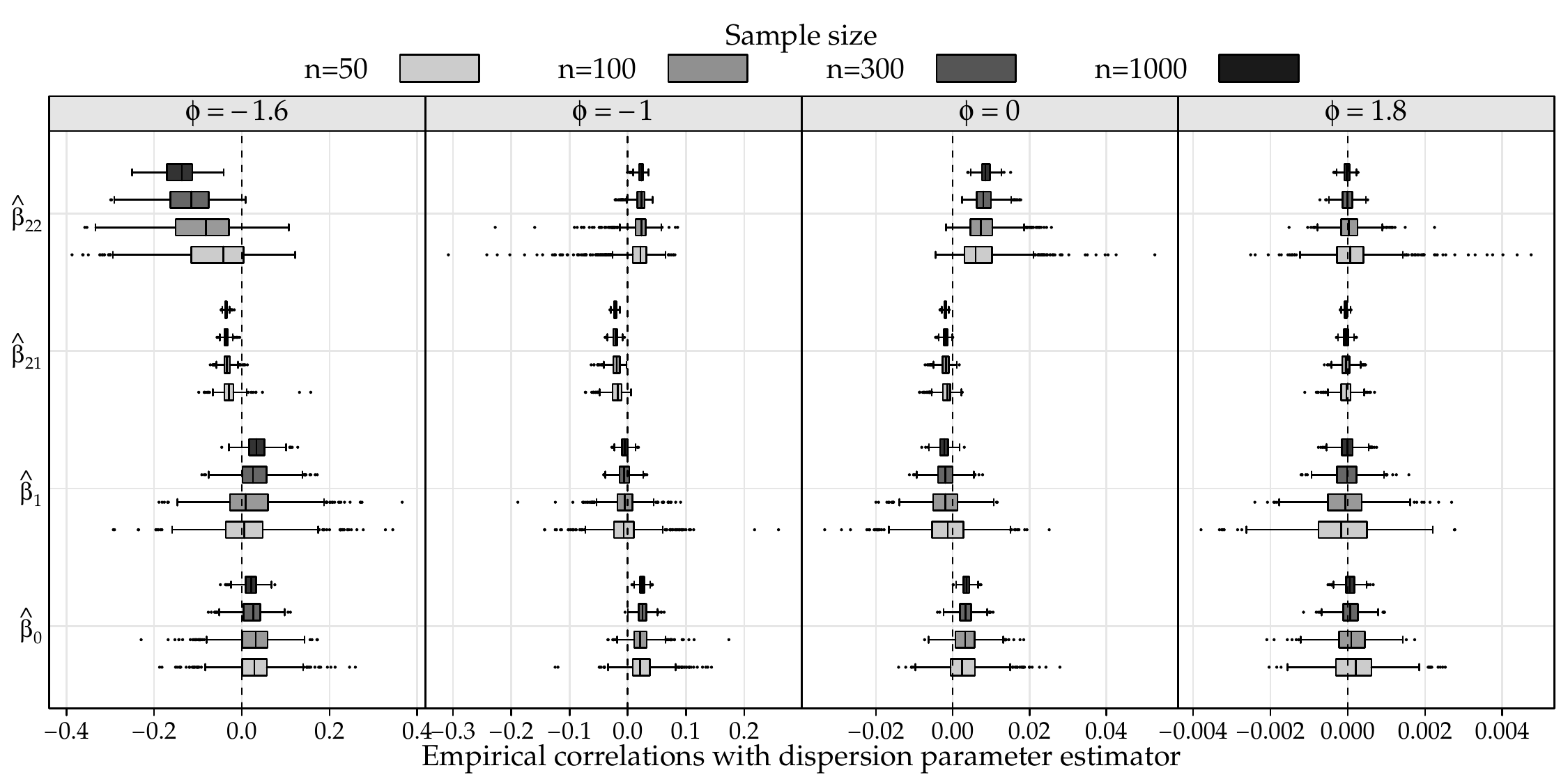} 

}

\caption[Empirical correlations between regression and dispersion parameters by different sample sizes and dispersion levels]{Empirical correlations between regression and dispersion parameters by different sample sizes and dispersion levels.}\label{fig:ortho-plot}
\end{figure}

\end{knitrout}

\begin{knitrout}
\definecolor{shadecolor}{rgb}{0.969, 0.969, 0.969}\color{fgcolor}\begin{figure}[!htp]

{\centering \includegraphics[width=1\textwidth]{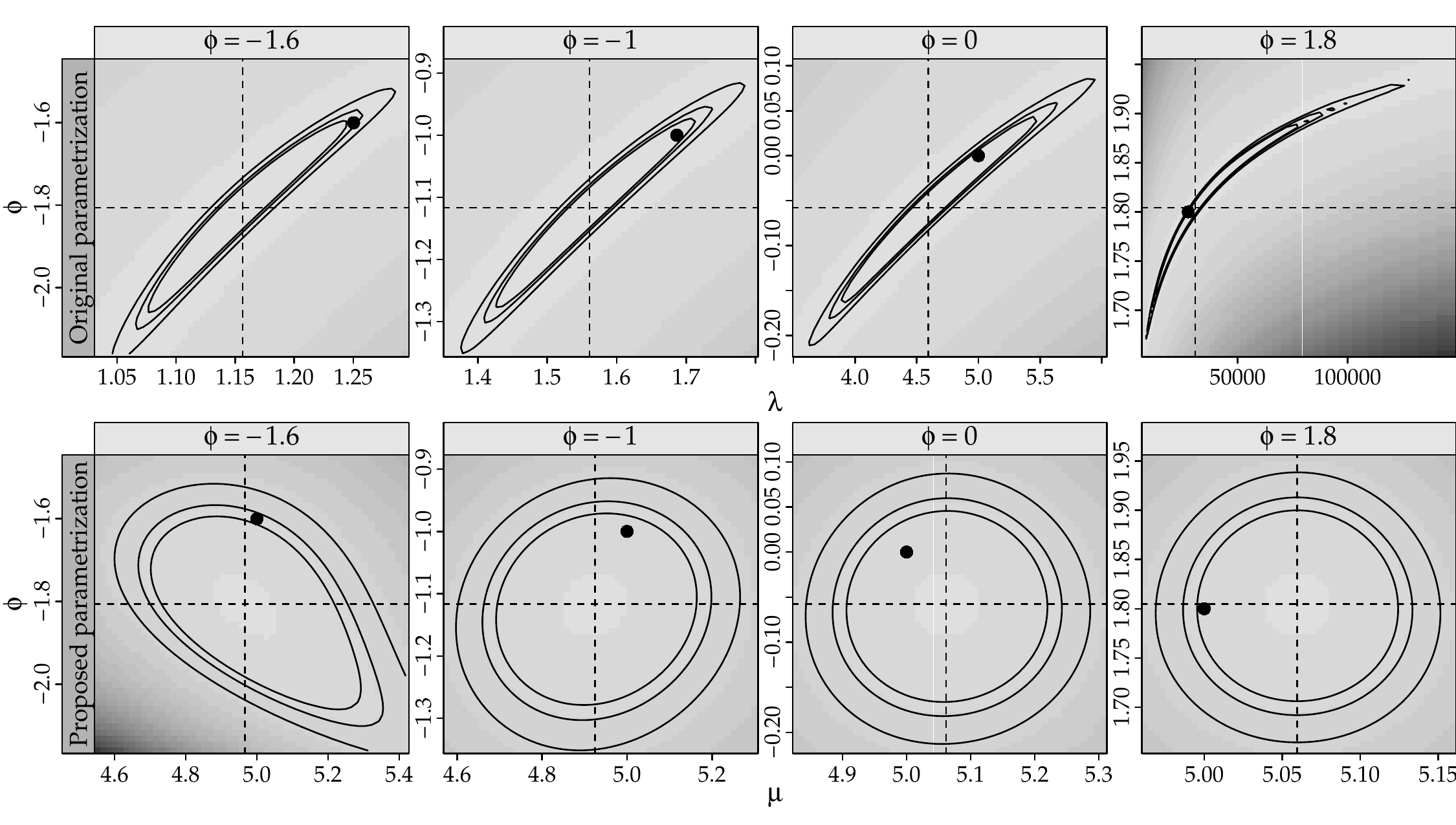} 

}

\caption[Deviance surface contour plots under original and proposed parametrization for four simulated data sets of size 1000 with different dispersion parameters]{Deviance surface contour plots under original and proposed parametrization for four simulated data sets of size 1000 with different dispersion parameters. The ellipses are confidence regions (90, 95 and 99\%), dotted lines are the maximum likelihood estimates, and points are the real parameters used in the simulation.}\label{fig:ortho-surf}
\end{figure}

\end{knitrout}

To illustrate the orthogonality, \autoref{fig:ortho-surf} displays contour plots of the deviance
surfaces for four simulated data set of size 1000, $\mu=5$ and different
values of the dispersion parameters. The shapes of the deviance function
show that the proposed parametrization is better for both computation
and asymptotic (normal-based) inference. Furthermore, it is interesting
to note that the deviance function shape under strong overdispersion
($\phi=-1.6$) is not as well behaved as the others;  this is due to the
difficulty of the distribution in dealing with strong overdispersion in
low counts (see dispersion index plot in the
\autoref{fig:indexes-plot}). This also explains the results of
Cov$(\hat{\beta}_{22}, \phi)$ in the first panel of
\autoref{fig:ortho-plot}, since $\beta_{22}$ is negative and associated
with low counts.

\section{Case studies}
\label{case-studies}

In this section, we report three illustrative examples of count data
analysis. We considered as alternative models for the analysis the
standard Poisson regression model, the COM-Poisson model in the two
forms (original and new parametrization) and the quasi-Poisson
regression model.  The data sets and \texttt{R} code for their analysis
are available as supplementary material.

\subsection{Artificial defoliation in cotton phenology}
\label{case-cotton}

This example relates to cotton plants (\textit{Gossypium hirsutum})
submitted to five levels of artificial defoliation (\texttt{des}) and
crossed with five growth stages (\texttt{est}). The main goal of this
study was to assess the effect of defoliation levels at different growth
stages of cotton plants on the cotton production, expressed by the
number of bolls produced. The study was conducted in a greenhouse and
the experimental design was completely randomized with five
replicates. This data set was analyzed by~\citet{Zeviani2014} using the
Gamma-Count distribution.

Following~\citet{Zeviani2014}, the linear predictor is given by
$$\log(\mu_{ij}) = \beta_0 + \beta_{1j} \texttt{def}_i + \beta_{2j}
\texttt{def}_i^2$$
where $\mu_{ij}$ is the expected number of cotton bolls for the $i$-th
defoliation level ($i=$ 1: 0\%, 2: 25\%, 3: 50\%, 4: 75\% e 5: 100\%)
and $j$-th growth stage ($j$ = 1: vegetative, 2: flower bud, 3: blossom,
4: boll, 5: boll open), i.e. we have a second order effect of
defoliation in each growth stage. The parameters estimates and
goodness-of-fit measures for the Poisson, COM-Poisson, COM-Poisson$_\mu$
and quasi-Poisson regression models are presented in
\autoref{tab:coef-cotton}.

\begin{table}[!ht]
\centering \small
\caption{Parameter estimates (Est) and ratio between estimate and
  standard error (SE) for the four model strategies for the analysis
  of the cotton experiment.}
\label{tab:coef-cotton}
\begin{tabular}{lrrrrrrrr}
  \toprule
  & \multicolumn{2}{c}{Poisson} &
    \multicolumn{2}{c}{COM-Poisson} &
    \multicolumn{2}{c}{COM-Poisson$_\mu$} &
    \multicolumn{2}{c}{Quasi-Poisson} \\
\cmidrule(lr){2-3} \cmidrule(lr){4-5} \cmidrule(lr){6-7} \cmidrule(lr){8-9}
 & Est & Est/SE & Est & Est/SE & Est & Est/SE & Est & Est/SE \\ 
  \midrule
$\phi\,,\,\sigma$ &  &  & 1.5847 & 12.4166 & 1.5817 & 12.3922 & 0.2410 &  \\ 
  $\beta_0$ & 2.1896 & 34.5724 & 10.8967 & 7.7594 & 2.1905 & 74.6397 & 2.1896 & 70.4205 \\ 
  $\beta_{11}$ & 0.4369 & 0.8473 & 2.0187 & 1.7696 & 0.4350 & 1.8194 & 0.4369 & 1.7260 \\ 
  $\beta_{12}$ & 0.2897 & 0.5706 & 1.3431 & 1.2109 & 0.2876 & 1.2227 & 0.2897 & 1.1622 \\ 
  $\beta_{13}$ & $-$1.2425 & $-$2.0581 & $-$5.7505 & $-$3.8858 & $-$1.2472 & $-$4.4202 & $-$1.2425 & $-$4.1921 \\ 
  $\beta_{14}$ & 0.3649 & 0.6449 & 1.5950 & 1.2975 & 0.3500 & 1.3280 & 0.3649 & 1.3135 \\ 
  $\beta_{15}$ & 0.0089 & 0.0178 & 0.0377 & 0.0346 & 0.0076 & 0.0324 & 0.0089 & 0.0362 \\ 
  $\beta_{21}$ & $-$0.8052 & $-$1.3790 & $-$3.7245 & $-$2.7754 & $-$0.8033 & $-$2.9613 & $-$0.8052 & $-$2.8089 \\ 
  $\beta_{22}$ & $-$0.4879 & $-$0.8613 & $-$2.2647 & $-$1.8051 & $-$0.4858 & $-$1.8499 & $-$0.4879 & $-$1.7544 \\ 
  $\beta_{23}$ & 0.6728 & 0.9892 & 3.1347 & 2.0837 & 0.6788 & 2.1349 & 0.6728 & 2.0149 \\ 
  $\beta_{24}$ & $-$1.3103 & $-$1.9477 & $-$5.8943 & $-$3.6567 & $-$1.2875 & $-$4.0951 & $-$1.3103 & $-$3.9672 \\ 
  $\beta_{25}$ & $-$0.0200 & $-$0.0361 & $-$0.0901 & $-$0.0755 & $-$0.0189 & $-$0.0740 & $-$0.0200 & $-$0.0736 \\ 
   \specialrule{0.01em}{0.3em}{0.3em} 
 LogLik & \multicolumn{2}{c}{$-255.803$} & \multicolumn{2}{c}{$-208.250$} & \multicolumn{2}{c}{$-208.398$} & \multicolumn{2}{c}{$  -$}\\
 AIC & \multicolumn{2}{c}{$533.606$} & \multicolumn{2}{c}{$440.500$} & \multicolumn{2}{c}{$440.795$} & \multicolumn{2}{c}{$  -$}\\
 BIC & \multicolumn{2}{c}{$564.718$} & \multicolumn{2}{c}{$474.440$} & \multicolumn{2}{c}{$474.735$} & \multicolumn{2}{c}{$  -$} \\
 \bottomrule
\end{tabular}
\end{table}

The results presented in \autoref{tab:coef-cotton} show that the
goodness-of-fit measures (log-likelihood, AIC and BIC) are quite similar
for the COM-Poisson and COM-Poisson$_\mu$ models.  It suggests that the
reparametrization does not change the model fit, as expected.  The
Poisson model is clearly unsuitable, being overly conservative.  The
difference in terms of log-likelihood value from the Poisson to the
COM-Poisson$_\mu$ model was $94.811$, which in turn
suggests the better fit of the COM-Poisson$_{\mu}$ model. A chi-square
test also supports this statement.  Finally, the estimated value of the
dispersion parameter $\\hat{\phi} = 1.582$ suggests
underdispersion.

Furthermore, results in \autoref{tab:coef-cotton} also show the
advantage of the COM-Poisson$_\mu$ model, since the estimates are quite
similar to the ones obtained by the Poisson model, whereas estimates
obtained from the COM-Poisson model in the original parametrization are
on a non interpretable scale.  The ratios between estimates and their
respective standard errors for the COM-Poisson models are very close to
ratios obtained by quasi-Poisson model. However, it is important to note
that COM-Poisson model is a full parametric approach, i.e. there is a
probability distribution associated to the counts. On the other hand,
the quasi-Poisson model is a specification based only on second-moment
assumptions.

\autoref{fig:pred-cotton} presents the observed and fitted values with
confidence intervals (95\%) as a function of the defoliation level for
each growth stage. The fitted values are the same for the Poisson and
COM-Poisson$_{\mu}$ models, however, the confidence intervals are larger
for the Poisson model because the equidispersion assumption.  The
results from the COM-Poisson$_{\mu}$ model are consistent with those
from the Gamma-Count model~\citep{Zeviani2014},
Poisson-Tweedie~\citep{Bonat2017} and the alternative parametrization of
the COM-Poisson distribution proposed by~\citet{Huang2017}.  In all
strategies the models indicated underdispersion and significant effects
of defoliation for the vegetative, blossom and boll growth stages.

\begin{knitrout}
\definecolor{shadecolor}{rgb}{0.969, 0.969, 0.969}\color{fgcolor}\begin{figure}[!htp]

{\centering \includegraphics[width=1\textwidth]{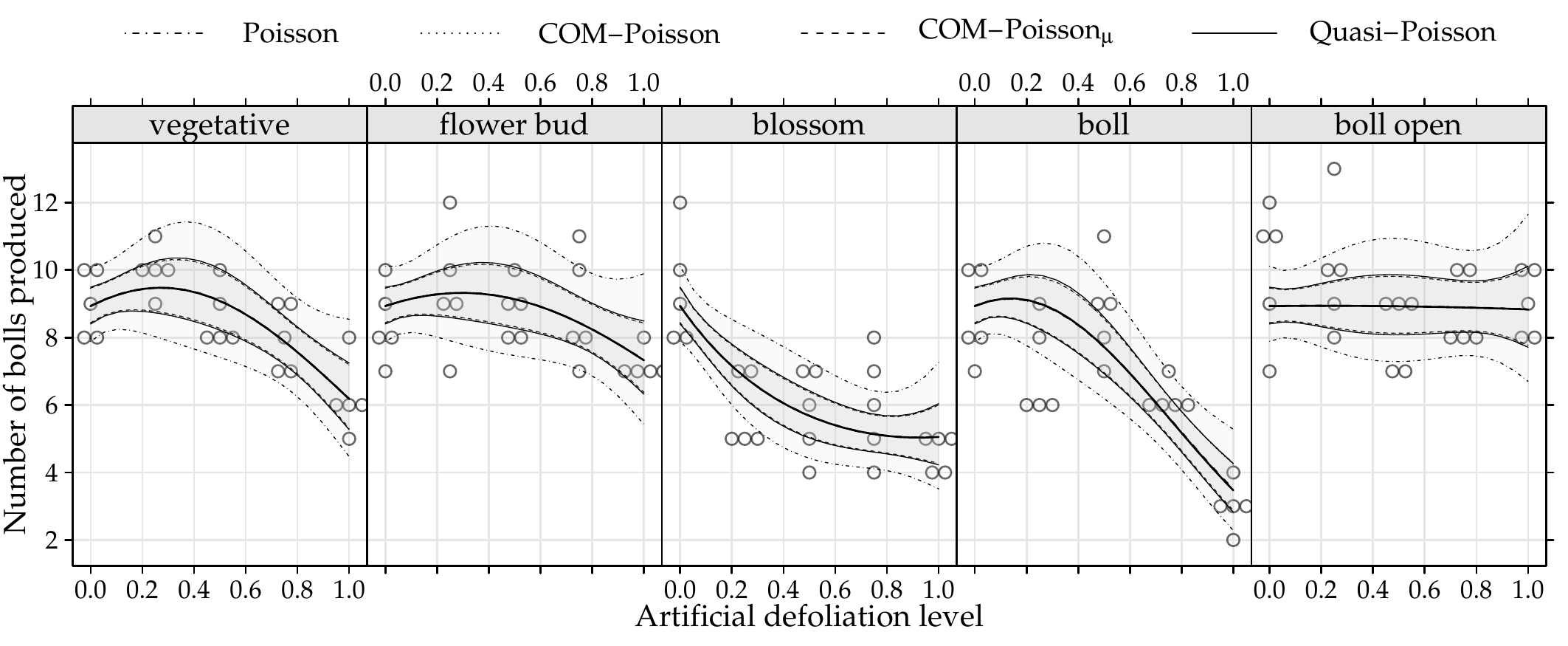} 

}

\caption[Scatterplots of the observed data and curves of fitted values with 95\% confidence intervals as functions of the defoliation level for each growth stage]{Scatterplots of the observed data and curves of fitted values with 95\% confidence intervals as functions of the defoliation level for each growth stage.}\label{fig:pred-cotton}
\end{figure}

\end{knitrout}

In order to assess the relation between $\bm{\mu}$ and $\phi$ in the
COM-Poisson$_\mu$ parametrization, \autoref{tab:corr-cotton} presents
the empirical correlations between the regression and dispersion
parameters, as computed by the asymptotic covariance matrix of the
estimators, i.e. the inverse of the observed information.  The
correlations are practically null considering the COM-Poisson$_\mu$. On
the other hand, for the original parametrization such correlations are
quite large, in particular for the parameter $\beta_0$ (due to effects
parametrization in the linear predictor).  This result explain the
better performance of the maximization algorithm in the new
parametrization.  It is important to note that the initial values for
the \texttt{BFGS} algorithm are provided by the Poisson model, then in
the COM-Poisson$_{\mu}$ model the initial values are practically the
maximum likelihood estimates and the effort of maximization is on the
dispersion parameter $\phi$ only. To compare the computational times on
the two parametrizations we repeat the fitting $50$ times. In
this case COM-Poisson$_\mu$ fit was, on average, $38$\% faster than the
original one.

\begin{table}[ht]
\centering
\caption{Empirical correlations between $\hat{\phi}$ and
  $\hat{\bm{\beta}}$ for the two parametrizations of COM-Poisson model
  fit to underdispersed data.}
\label{tab:corr-cotton}
\begingroup\small
\begin{tabular}{lrrrrrr}
  \toprule
 & $\hat{\beta}_0$ & $\hat{\beta}_{11}$ & $\hat{\beta}_{12}$ & $\hat{\beta}_{13}$ & $\hat{\beta}_{14}$ & $\hat{\beta}_{15}$ \\ 
  \midrule
COM-Poisson & 0.9952 & 0.2229 & 0.1526 & $-$0.4895 & 0.1614 & 0.0043 \\ 
  COM-Poisson$_\mu$ & 0.0005 & $-$0.0002 & $-$0.0002 & $-$0.0007 & $-$0.0015 & $-$0.0002 \\ 
  \midrule
 & $\hat{\beta}_{21}$ & $\hat{\beta}_{22}$ & $\hat{\beta}_{23}$ & $\hat{\beta}_{24}$ & $\hat{\beta}_{25}$ & $ $ \\ 
  \midrule
COM-Poisson & $-$0.3496 & $-$0.2276 & 0.2629 & $-$0.4578 & $-$0.0095 &  \\ 
  COM-Poisson$_\mu$ & 0.0001 & 0.0002 & 0.0006 & 0.0018 & 0.0001 &  \\ 
   \bottomrule

\end{tabular}
\endgroup
\end{table}

\subsection{Soil moisture and potassium doses on soybean culture}
\label{case-soybean}

The second example is a $5\times 3$ factorial experiment in a randomized
complete block design. The aim of this study was to evaluate the effects
of potassium doses (\texttt{K}) applied to soil (0, 0.3, 0.6, 1.2 and
1.8 $\times$ 100mg dm$^{-3}$) and soil moisture (\texttt{umid}) levels
(37.5, 50, 62.5\%) on  soybean (\emph{Glicine Max}) production. The
experiment was carried out in a greenhouse, in pots with two plants, and
the count variable measured was the number of bean seeds per
pot~\citep{Serafim2012}.  \autoref{fig:desc-soy} (left) shows the number
of bean seeds recorded for each combination of potassium dose and moisture
level, it is important to note the indication of a quadratic effect of
the potassium levels as shown by smoothing curves. Most points in the
sample variance \textit{versus} sample means dispersion diagram (right)
are above the identity line, suggesting overdispersion (block effect not
yet removed).

\begin{knitrout}
\definecolor{shadecolor}{rgb}{0.969, 0.969, 0.969}\color{fgcolor}\begin{figure}[!ht]

{\centering \includegraphics[width=1\textwidth]{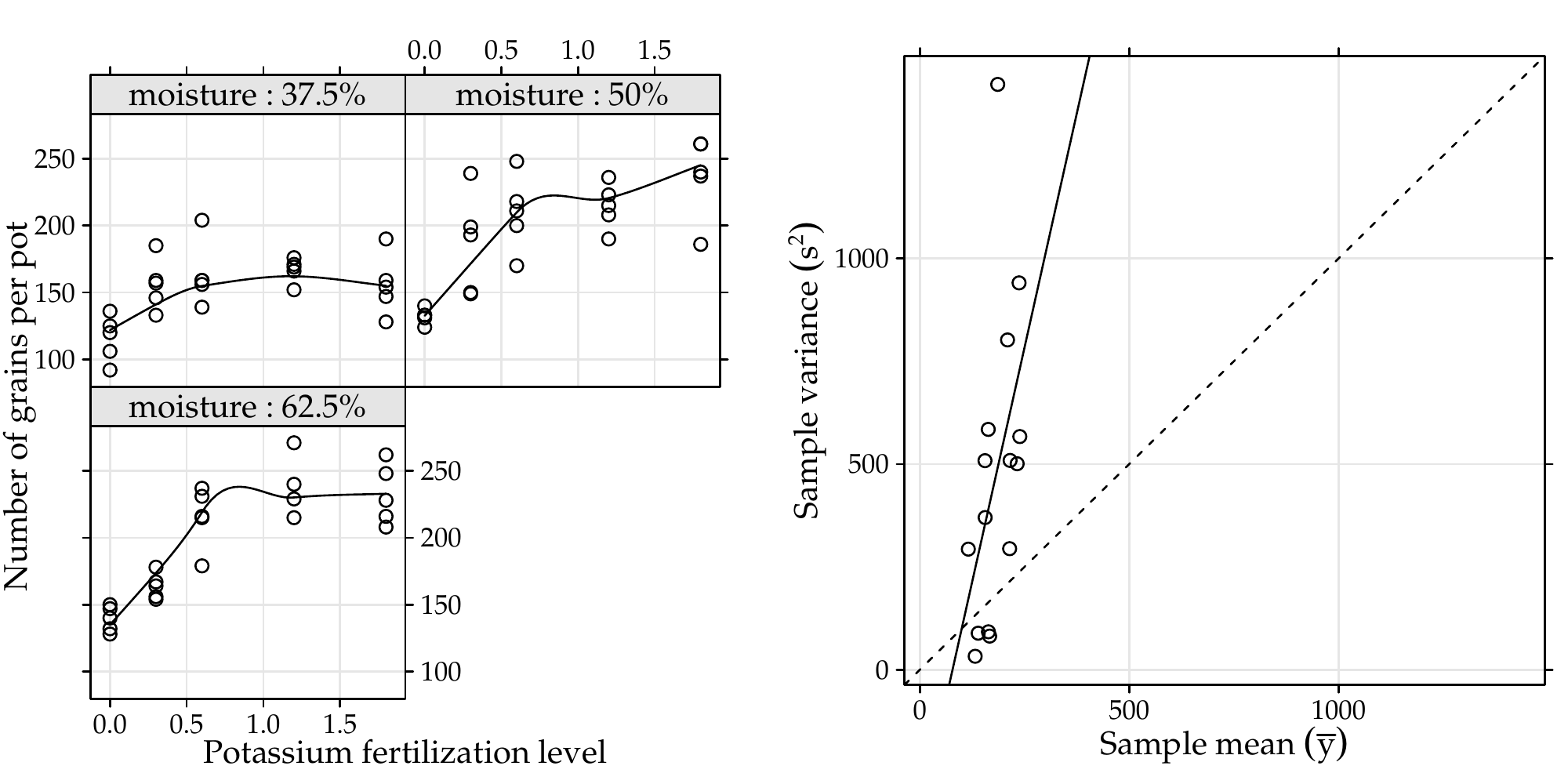} 

}

\caption[Number of bean seeds per pot for each potassium dose and moisture level (left) and sample mean against sample variance of the five replicates for each experimental treatment (right)]{Number of bean seeds per pot for each potassium dose and moisture level (left) and sample mean against sample variance of the five replicates for each experimental treatment (right). Solid lines are the smoothing curves on the left and the least of squares curve on the right.}\label{fig:desc-soy}
\end{figure}

\end{knitrout}

For the analysis of this data set based on the descriptive analysis
(\autoref{fig:desc-soy}), we proposed the following linear predictor
$$
\log(\mu_{ijk}) = \beta_0 + \gamma_i + \tau_j +
  \beta_{1}\texttt{K}_k + \beta_{2}\texttt{K}_k^2 +
  \beta_{3j}\texttt{K}_k
$$
where $i=$1: block II, 2: block III, 3: block IV e 4: block V; $j=$1:
50\% e 2: 62.5\%; and $k=$1: 0.0, 2: 0.3, 3: 0.6, 4: 1.2, 5: 1.8 100mg
dm$^{-3}$, where $\gamma_i$ is the effect of $i$-th block ($i=$1: block
II, 2: block III, 3: block IV and 4: block V), $\tau_j$ is the effect of
$j$-th moisture level ($j=$1: 50\% and 2: 62.5\%) and $\beta_{3j}$ is
interaction of the first order potassium effect (\texttt{K}) for the
$j$-th moisture level (\texttt{umid}). \autoref{tab:coef-soy} presents
the estimates, ratio between estimate and standard error and
goodness-of-fit measures for the alternative models.

The results in \autoref{tab:coef-soy} show that the two parametrization
of COM-Poisson model presented very similar goodness-of-fit measures and
better fit than the Poisson model. The difference between the
log-likelihoods of the Poisson and COM-Poisson models was
$29.697$, indicating that $\phi$ is significantly
different from zero. The estimate of $\phi$ ($-0.782$) indicates
overdispersion, corroborating the descriptive analysis. Concerning to
the regression parameters, the similarities between the models are
analogous to the previous section. Both models indicate effects of
block, potassium dose and moisture level, however the Poisson model
indicates the effects with greater significance, because it does not
take account of the extra variability.

\begin{table}[!ht]
\centering \small
\caption{Parameter estimates (Est) and ratio between estimate and
  standard error (SE) for the four model strategies for the analysis of
  the soybean experiment.}
\label{tab:coef-soy}
\begin{tabular}{lrrrrrrrr}
  \toprule
  & \multicolumn{2}{c}{Poisson} &
    \multicolumn{2}{c}{COM-Poisson} &
    \multicolumn{2}{c}{COM-Poisson$_\mu$} &
    \multicolumn{2}{c}{Quasi-Poisson} \\
\cmidrule(lr){2-3} \cmidrule(lr){4-5} \cmidrule(lr){6-7} \cmidrule(lr){8-9}
 & Est & Est/SE & Est & Est/SE & Est & Est/SE & Est & Est/SE \\ 
  \midrule
$\phi\,,\,\sigma$ &  &  & $-$0.7785 & $-$4.7208 & $-$0.7821 & $-$4.7371 & 2.6151 &  \\ 
  $\beta_0$ & 4.8666 & 144.2886 & 2.2320 & 6.0415 & 4.8666 & 97.7808 & 4.8666 & 89.2254 \\ 
  $\gamma_{1}$ & $-$0.0194 & $-$0.7302 & $-$0.0089 & $-$0.4939 & $-$0.0194 & $-$0.4950 & $-$0.0194 & $-$0.4516 \\ 
  $\gamma_{2}$ & $-$0.0366 & $-$1.3733 & $-$0.0169 & $-$0.9212 & $-$0.0366 & $-$0.9306 & $-$0.0366 & $-$0.8492 \\ 
  $\gamma_{3}$ & $-$0.1056 & $-$3.8890 & $-$0.0486 & $-$2.4223 & $-$0.1056 & $-$2.6338 & $-$0.1056 & $-$2.4049 \\ 
  $\gamma_{4}$ & $-$0.0916 & $-$3.2997 & $-$0.0422 & $-$2.1020 & $-$0.0917 & $-$2.2366 & $-$0.0916 & $-$2.0405 \\ 
  $\tau_{1}$ & 0.1320 & 3.6471 & 0.0609 & 2.2949 & 0.1320 & 2.4715 & 0.1320 & 2.2553 \\ 
  $\tau_{2}$ & 0.1243 & 3.4319 & 0.0573 & 2.1772 & 0.1243 & 2.3258 & 0.1243 & 2.1222 \\ 
  $\beta_1$ & 0.6160 & 11.0139 & 0.2839 & 4.7291 & 0.6161 & 7.4640 & 0.6160 & 6.8108 \\ 
  $\beta_2$ & $-$0.2759 & $-$10.2501 & $-$0.1272 & $-$4.5890 & $-$0.2760 & $-$6.9458 & $-$0.2759 & $-$6.3385 \\ 
  $\beta_{31}$ & 0.1456 & 4.2680 & 0.0670 & 2.6138 & 0.1456 & 2.8922 & 0.1456 & 2.6392 \\ 
  $\beta_{32}$ & 0.1648 & 4.8294 & 0.0759 & 2.8843 & 0.1648 & 3.2723 & 0.1648 & 2.9864 \\ 
   \specialrule{0.01em}{0.3em}{0.3em} 
 LogLik & \multicolumn{2}{c}{$-340.082$} & \multicolumn{2}{c}{$-325.241$} & \multicolumn{2}{c}{$-325.233$} & \multicolumn{2}{c}{$  -$}\\
 AIC & \multicolumn{2}{c}{$702.164$} & \multicolumn{2}{c}{$674.482$} & \multicolumn{2}{c}{$674.467$} & \multicolumn{2}{c}{$  -$}\\
 BIC & \multicolumn{2}{c}{$727.508$} & \multicolumn{2}{c}{$702.130$} & \multicolumn{2}{c}{$702.116$} & \multicolumn{2}{c}{$  -$} \\
 \bottomrule
\end{tabular}
\end{table}

The infinite sum $Z(\mu, \phi)$ in the cases of overdispersed count data
requires a larger upper bound to reach convergence. Thus, in these cases
the computation of the log-likelihood function is computationally
expensive.  For the data set considered, the upper bound was fixed at
$700$.  The \texttt{BFGS} algorithm evaluated the log-likelihood
function $264$ and $20$ times to reach
convergence, when using the original and new parametrization of the
COM-Poisson distribution, respectively.  In terms of computational time,
for 50 repetitions of fit, the proposed reparametrization was on average
$110\%$ faster than the original one.  Probably, it is due to the better
behaviour of the log-likelihood function as well as better initial
values obtained from the Poisson fit. In \autoref{tab:corr-soy}, we
present the empirical correlation between the regression and dispersion
parameter estimates. The correlations are close to zero for the
COM-Poisson$_\mu$ model, indicating the empirical orthogonality between
$\mu$ and $\phi$.

\begin{table}[ht]
\centering
\caption{Empirical correlations between $\hat{\phi}$ and
  $\hat{\bm{\beta}}$ for the two parametrizations of COM-Poisson model
  fit to overdispersed data.}
\label{tab:corr-soy}
\begingroup\small
\begin{tabular}{lrrrrrrr}
  \toprule
 & $\hat{\beta}_0$ & $\hat{\beta}_{11}$ & $\hat{\beta}_{12}$ & $\hat{\beta}_{13}$ & $\hat{\beta}_{14}$ & $\hat{\beta}_{15}$ & $\hat{\beta}_{21}$ \\ 
  \midrule
COM-Poisson & 0.9952 & 0.2229 & 0.1526 & $-$0.4895 & 0.1614 & 0.0043 & $-$0.3496 \\ 
  COM-Poisson$_\mu$ & 0.0005 & $-$0.0002 & $-$0.0002 & $-$0.0007 & $-$0.0015 & $-$0.0002 & 0.0001 \\ 
  \midrule
 & $\hat{\beta}_{22}$ & $\hat{\beta}_{23}$ & $\hat{\beta}_{24}$ & $\hat{\beta}_{25}$ & $ $ & $ $ & $ $ \\ 
  \midrule
COM-Poisson & $-$0.2276 & 0.2629 & $-$0.4578 & $-$0.0095 &  &  &  \\ 
  COM-Poisson$_\mu$ & 0.0002 & 0.0006 & 0.0018 & 0.0001 &  &  &  \\ 
   \bottomrule

\end{tabular}
\endgroup
\end{table}

The observed and fitted counts for each humidity level with confidence
intervals are shown in \autoref{fig:pred-soy}. The fitted values are
identical for the Poisson and COM-Poisson$_{\mu}$ models, leading to the
same conclusions. On the other hand, confidence intervals for the
Poisson model are smaller than the ones from the COM-Poisson$_{\mu}$,
due to the equidispersion assumption underlying the Poisson model. The
confidence intervals from the COM-Poisson$_{\mu}$ and quasi-Poisson
models are really similar, which in turn shows the already highlighted
similarity between these approaches, however only the COM-Poisson
model$_{\mu}$ corresponds to a fully specified probability model.

\begin{knitrout}
\definecolor{shadecolor}{rgb}{0.969, 0.969, 0.969}\color{fgcolor}\begin{figure}[!ht]

{\centering \includegraphics[width=1\textwidth]{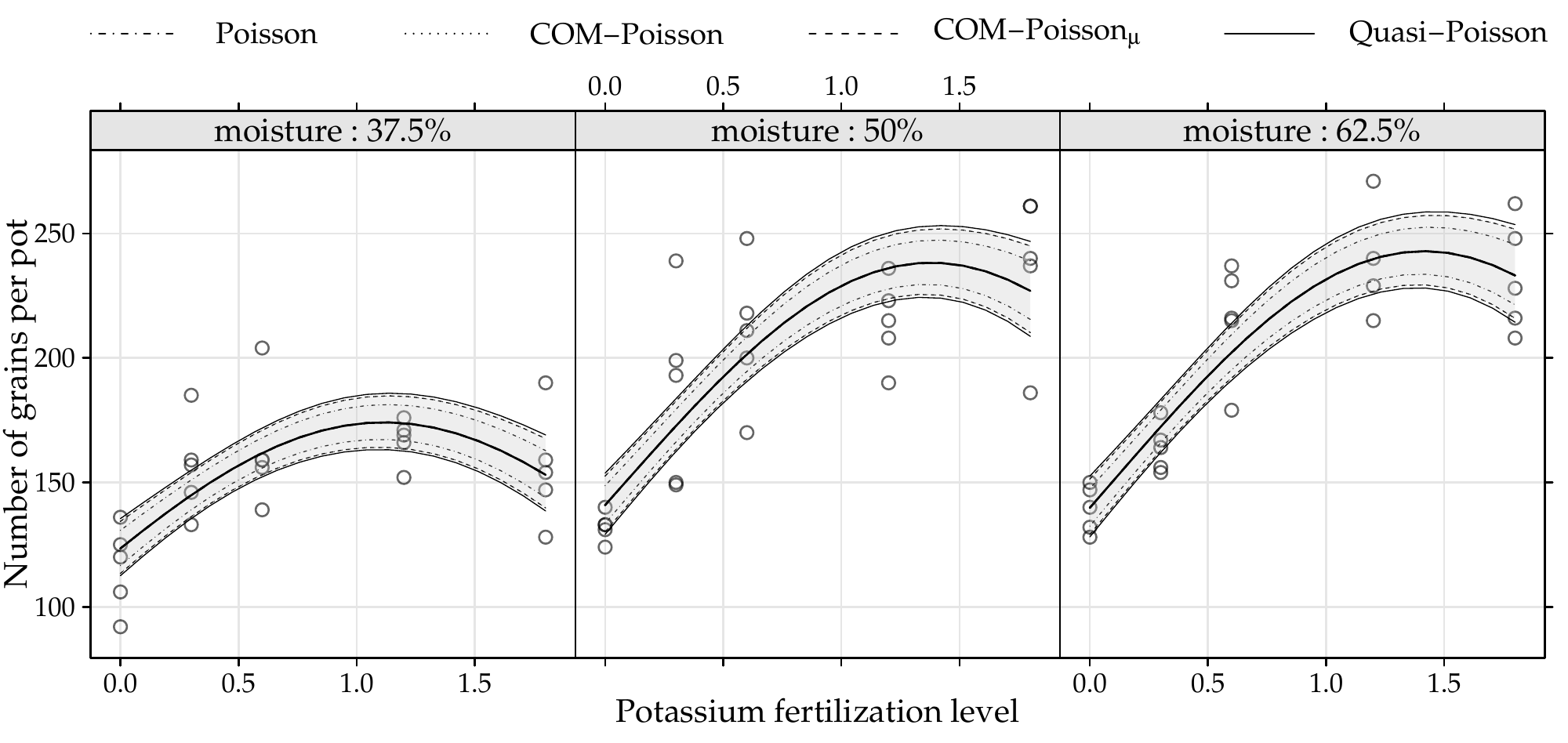} 

}

\caption[Dispersion diagrams of been seeds counts as function of potassium doses and humidity levels with fitted curves and confidence intervals (95\%)]{Dispersion diagrams of been seeds counts as function of potassium doses and humidity levels with fitted curves and confidence intervals (95\%).}\label{fig:pred-soy}
\end{figure}

\end{knitrout}

\subsection{Assessing toxicity of nitrofen in aquatic systems}

Nitrofen is a herbicide that was used extensively for the control of
broad-leaved and grass weeds in cereals and rice. Although it is
relatively non-toxic to adult mammals, nitrofen is a significant
tetragen and mutagen. It is also acutely toxic and reproductively toxic
to cladoceran zooplankton. Nitrofen is no longer in commercial use in
the U.S., having been the first pesticide to be withdrawn due to
tetragenic effects~\citep{Bailer1994}.

The data set comes from an experiment to measure the reproductive
toxicity of the herbicide, nitrofen, on a species of zooplankton
(\textit{Ceriodaphnia dubia}). Fifty animals were randomized into
batches of ten and each batch was put in a solution with a measured
concentration of nitrofen ($0, 0.8, 1.6, 2.35$ and $3.10$
$\mu$g$/10^2$litre) (\texttt{dose}). Then the number of live offspring
was recorded.

For this data set we consider three models with linear predictors,
\begin{center}
\begin{minipage}{12cm}
Linear: $\log(\mu_i) = \beta_0 + \beta_1 \texttt{dose}_i$\\
Quadratic: $\log(\mu_i) = \beta_0 + \beta_1 \texttt{dose}_i +
             \beta_2 \texttt{dose}_i^2$\\
Cubic: $\log(\mu_i) = \beta_0 + \beta_1 \texttt{dose}_i +
             \beta_2 \texttt{dose}_i^2 + \beta_3 \texttt{dose}_i^3$.
\end{minipage}
\end{center}

\begin{table}[!ht]
\centering \small
\caption{Model fit measures and comparisons between predictors and
  models fitted to the nitrofen data.}
\label{tab:anova-ovos}
\begin{tabularx}{\textwidth}{lCCCCCrC}
  \toprule
 Poisson & np & $\ell$ & AIC & 2(diff $\ell$) & diff np & P($>\rchi^2$) & \\
 \midrule
 Preditor 1 & 2 & $-$180.667 & 365.335 &  &  &  &  \\ 
  Preditor 2 & 3 & $-$147.008 & 300.016 & 67.319 & 1 & 2.31E$-$16 &  \\ 
  Preditor 3 & 4 & $-$144.090 & 296.180 & 5.835 & 1 & 1.57E$-$02 &  \\ 
  
\specialrule{0em}{0.5em}{0em} 
  COM-Poisson & np & $\ell$ & AIC & 2(diff $\ell$) & diff np &
  P($>\rchi^2$) & $\hat{\phi}$ \\
  \midrule
 Preditor 1 & 3 & $-$167.954 & 341.908 &  &  &  & $-$0.893 \\ 
  Preditor 2 & 4 & $-$146.964 & 301.929 & 41.980 & 1 & 9.22E$-$11 & $-$0.059 \\ 
  Preditor 3 & 5 & $-$144.064 & 298.129 & 5.800 & 1 & 1.60E$-$02 & 0.048 \\ 
  
\specialrule{0em}{0.5em}{0em} 
  COM-Poisson$_\mu$ & np & $\ell$ & AIC & 2(diff $\ell$) & diff np &
  P($>\rchi^2$) & $\hat{\phi}$ \\
  \midrule
 Preditor 1 & 3 & $-$167.652 & 341.305 &  &  &  & $-$0.905 \\ 
  Preditor 2 & 4 & $-$146.950 & 301.900 & 41.405 & 1 & 1.24E$-$10 & $-$0.069 \\ 
  Preditor 3 & 5 & $-$144.064 & 298.127 & 5.773 & 1 & 1.63E$-$02 & 0.047 \\ 
  
\specialrule{0em}{0.5em}{0em} 
  Quasi-Poisson & np & QDev & AIC & F & diff np & P($>F$) & $\hat{\sigma}$ \\
  \midrule
 Preditor 1 & 3 & 123.929 &  &  &  &  & 2.262 \\ 
  Preditor 2 & 4 & 56.610 &  & 60.840 & 1 & 5.07E$-$10 & 1.106 \\ 
  Preditor 3 & 5 & 50.774 &  & 5.659 & 1 & 2.16E$-$02 & 1.031 \\ 
  
 \bottomrule
\end{tabularx}

\footnotesize \raggedright np, number of parameters; diff $\ell$,
difference in log-likelihoods; QDev, quasi-deviance, F, F statistics
based on quasi-deviances; diff np, difference in np.
\end{table}

\autoref{tab:anova-ovos} summarizes the results of the fitted models and
likelihood ratio tests comparing the sequence of predictors. All models
indicate the significance of the cubic effect of the nitrofen
concentration.  Considering this predictor, there is an evidence of
equidispersed counts, the $\phi$ estimate of the COM-Poisson is close to
zero and $\sigma$ of quasi-Poisson is close to one.  It is interesting
to note that if we omit the high order effects the models show evidence
of overdispersion. This exemplifies the discussion on the causes of
overdispersion made in \autoref{introduction}.  We can also note that
the quasi-Poisson approach, although robust to equidispersion
assumption, shows higher descriptive levels ($p$-values) than parametric
models, that is, the tests under parametric models are more powerful
than the ones under the quasi-Poisson model in the equidispersed case.

In \autoref{tab:coef-ovos}, we present the estimates of the regression
parameters considering the cubic dose model. The interpretations are
similar to others cases studies, however, in this case the Poisson model
is also suitable for indicating the significance of the covariate
effects.  In addition, note that the parameter estimates of the original
COM-Poisson model are comparable to the others models. This occurs
because we are in the particular case, where $\phi = 0$, which implies
$\lambda = \mu$.

\begin{table}[!ht]
\centering
\caption{Parameter estimates (Est) and ratio between estimate and
  standard error (SE) for the four model strategies for the analysis of
  the nitrofen experiment.}
\label{tab:coef-ovos}
\begin{tabular}{lrrrrrrrr}
  \toprule
  & \multicolumn{2}{c}{Poisson} &
    \multicolumn{2}{c}{COM-Poisson} &
    \multicolumn{2}{c}{COM-Poisson$_\mu$} &
    \multicolumn{2}{c}{Quasi-Poisson} \\
\cmidrule(lr){2-3} \cmidrule(lr){4-5} \cmidrule(lr){6-7} \cmidrule(lr){8-9}
 & Est & Est/SE & Est & Est/SE & Est & Est/SE & Est & Est/SE \\ 
  \midrule
$\beta_{0}$ & 3.4767 & 62.8167 & 3.6494 & 4.8499 & 3.4769 & 64.3083 & 3.4767 & 61.8599 \\ 
  $\beta_{1}$ & $-$0.0860 & $-$0.4328 & $-$0.0914 & $-$0.4475 & $-$0.0879 & $-$0.4523 & $-$0.0860 & $-$0.4262 \\ 
  $\beta_{2}$ & 0.1529 & 0.8634 & 0.1612 & 0.8783 & 0.1547 & 0.8938 & 0.1529 & 0.8502 \\ 
  $\beta_{3}$ & $-$0.0972 & $-$2.3978 & $-$0.1021 & $-$2.2294 & $-$0.0976 & $-$2.4635 & $-$0.0972 & $-$2.3612 \\ 
   \bottomrule

\end{tabular}
\end{table}

\autoref{fig:pred-ovos} shows the number of live off-spring observed and
fitted curves along with confidence intervals for all model strategies
adopted.  The fitted values and confidence intervals are identical and
have a complete overlap. It shows that the estimation of the extra
dispersion parameter does not affect the estimation of the regression
coefficients in the case of equidispersed counts.

\begin{knitrout}
\definecolor{shadecolor}{rgb}{0.969, 0.969, 0.969}\color{fgcolor}\begin{figure}[!htp]

{\centering \includegraphics[width=0.7\textwidth]{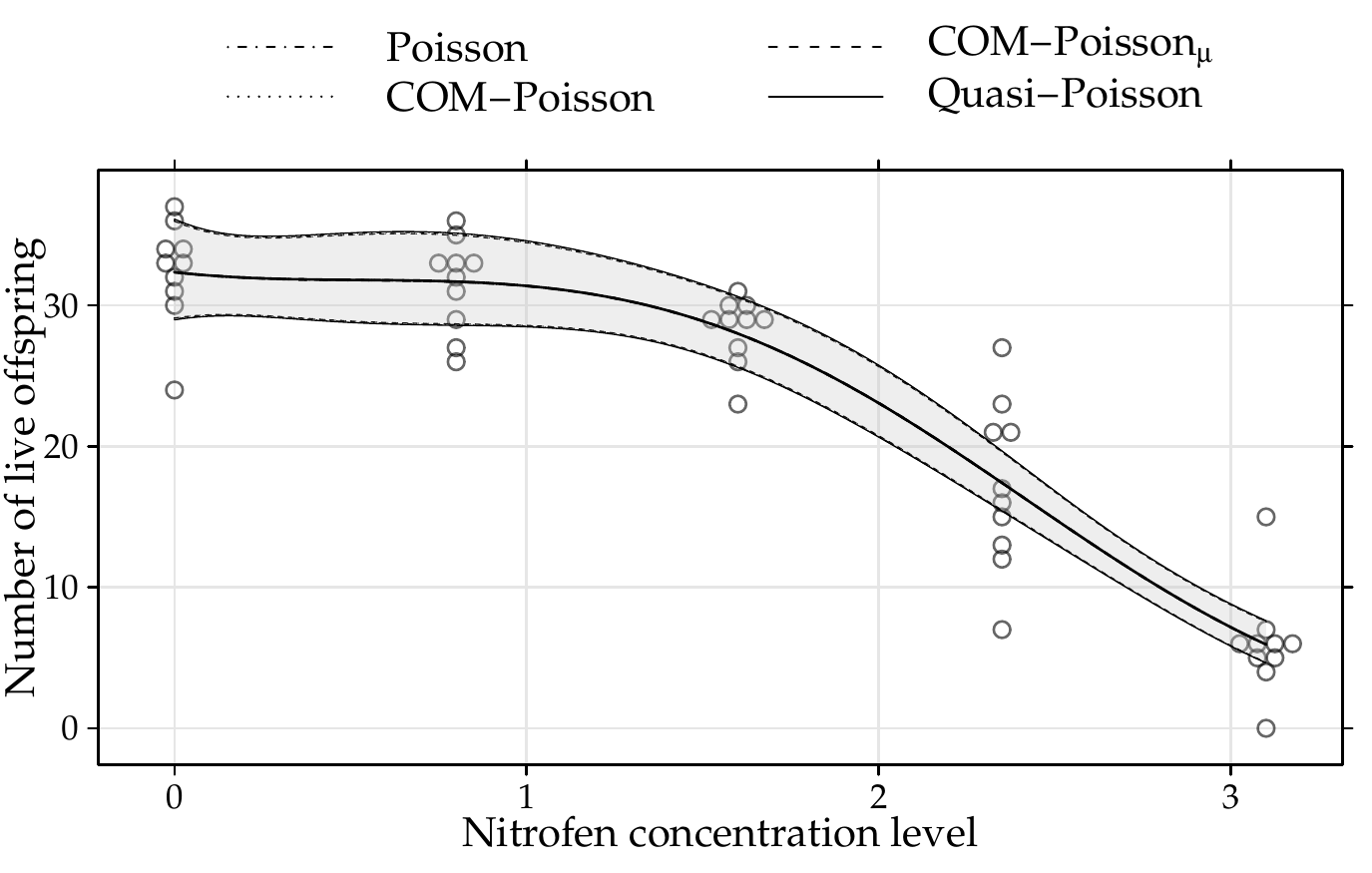} 

}

\caption[Number of live offsprings observed for each nitrofen concentration level with fitted curves and 95\% confidence intervals]{Number of live offsprings observed for each nitrofen concentration level with fitted curves and 95\% confidence intervals.}\label{fig:pred-ovos}
\end{figure}

\end{knitrout}

Finally, in \autoref{tab:corr-ovos} we present the empirical
correlations between the regression and dispersion parameter estimates.
The results show that even in the special case ($\phi=0$), the empirical
correlations for the original COM-Poisson model are not zero.  For the
reparametrized model, as discussed in the previous sections, the
correlations are practically null. The computational times for fifty
repetitions of fit are similar. The average time to fit for the
COM-Poisson$_\mu$ and COM-Poisson models is 1.19 and 1.09 seconds,
respectively.

\begin{table}[ht]
\centering
\caption{Empirical correlations between $\hat{\phi}$ and $\hat{\bm{\beta}}$ for the two parametrizations of COM-Poisson model fit to equidispersed data.} 
\label{tab:corr-ovos}
\begingroup\small
\begin{tabular}{lrrrr}
  \toprule
 & $\hat{\beta}_{0}$ & $\hat{\beta}_{1}$ & $\hat{\beta}_{2}$ & $\hat{\beta}_{3}$ \\ 
  \midrule
COM-Poisson & 0.9972 & $-$0.0771 & 0.1562 & $-$0.4223 \\ 
  COM-Poisson$_\mu$ & $-$0.0003 & 0.0023 & $-$0.0029 & 0.0033 \\ 
   \bottomrule
\end{tabular}
\endgroup
\end{table}

\section{Concluding remarks}
\label{conclusion}

In this paper, we presented and characterized a novel parametrization of
the COM-Poisson distribution and the associated regression model. The
novel parametrization was based on a simple asymptotic approximation for
the expectation and variance of the COM-Poisson distribution. The main
advantage of the proposed reparametrization is the simple interpretation
of the regression coefficients in terms of the expectation of the
response variable as usual in the generalized linear models
context. Thus, it is possible to compare the results of the COM-Poisson
model with the ones from standard approaches as the Poisson and
quasi-Poisson regression models. Furthermore, in the novel
parametrization the COM-Poisson distribution is indexed by the
expectation $\mu$ and an extra dispersion parameter $\phi$ which our
data analysis suggest being orthogonal. This is similar to
\Citeauthor{Huang2017}'s (\citeyear{Huang2017}) parametrization but is
simpler, because the $\mu$ is obtained from simple algebra.

We evaluated the accuracy of the asymptotic approximations for the
expectation and variance of the COM-Poisson distribution by considering
quadratic approximation errors.  The results showed that the
approximations are accurate for a large part of the parameter space,
which in turn support our reparametrization. Moreover, we discuss the
properties and flexibility of the distribution to deal with count data
although dispersion, zero-inflation and heavy-tail indexes. We carried
out a simulation study to assess the properties of the reparametrized
COM-Poisson model to deal with different levels of dispersion as well as
the properties of the maximum likelihood estimators. The results of our
simulation study suggested that the maximum likelihood estimators of the
regression and dispersion parameters are unbiased and consistent.  The
empirical coverage rates of the confidence intervals computed based on
the asymptotic distribution of the maximum likelihood estimators are
close to the nominal level for sample size greater than $100$.  The
worst scenario is when we have small sample sizes and strong
overdispersed counts.  In general, we recommend the use of the
asymptotic confidence intervals for computational simplicity.

The data analyses have shown that the COM-Poisson regression model is a
suitable choice to deal with dispersed count data.  The observed
empirical correlation between the regression and dispersion parameter
estimators suggest orthogonality between $\mu$ and $\phi$ in
COM-Poisson$_\mu$ distribution. Thus, the computational procedure based
on the proposed reparametrization is faster than in the original
parametrization.

In general, the results presented by the reparametrized COM-Poisson
models were satisfactory and comparable to the conventional approaches.
Therefore, its use in the analysis of count data is encouraged. The
computational routines for fitting the original and reparametrized
COM-Poisson regression models are available in the supplementary
material\textsuperscript{\ref{papercompanion}}.

There are many possible extensions to the model discussed in this paper,
including simulation studies to assess the model robustness against
model misspecification and to assess the theoretical approximations for
$Z(\lambda, \nu)$ (or $Z(\mu,\phi))$. Another simple extension of the
proposed model is to model both $\mu$ and $\phi$ parameters as functions
of covariate in a double generalized linear models framework.  Finally,
the reparametrized version of the COM-Poisson model also encourages the
specification of generalized linear mixed models using this
distribution.

\bibliography{references.bib}

\end{document}